\documentclass{jaa}
\usepackage[round]{natbib}
\usepackage{hyperref}
\hypersetup{colorlinks,citecolor=blue,urlcolor=blue,hypertexnames=true}
\usepackage{graphicx}

\begin{document}\sloppy

\title{Jets in radio galaxies and quasars: an observational perspective \\}


\author{D. J. Saikia\textsuperscript{1,2*}} 
\affilOne{\textsuperscript{1}Inter-University Centre for Astronomy and Astrophysics, Savitribai Phule Pune University Campus, Ganeshkhind, Pune 411007, India.\\}
\affilTwo{\textsuperscript{2}Department of Physics, Tezpur University, Napaam, Tezpur 784028, India.}


\twocolumn[{

\maketitle

\corres{dhrubasaikia@iucaa.in, dhrubasaikia.tifr.ccsu@gmail.com}

\msinfo{xxxx}{xxxx}

\begin{abstract}
This article gives a brief historical introduction and reviews our current understanding of jets in radio galaxies and quasars from an observational perspective, with an emphasis on observations at radio wavelengths. Recent results on the Fanaroff-Riley (FR) classification scheme, and the nature of radio structures and jets in the FR classes as well as in high-excitation and low-excitation radio galaxies are summarized. The collimation and propagation of jets from nuclear sub-pc to hundreds of kpc scales from both observational and theoretical work have been discussed. The jets exhibit evidence of interaction with a clumpy interstellar medium, especially in young radio sources, and could trigger both star formation as well as suppress star formation depending on the physical conditions. Observational evidence for such interactions and jet feedback which have profound implications in our understanding of galaxy evolution have been presented. Recurrent jet activity which has been seen over a wide range of projected linear size and time scales has been discussed. This review article concludes with a brief discussion of unresolved questions on jets which new telescopes should help address. \end{abstract}


}]


\doinum{12.3456/s78910-011-012-3}
\artcitid{\#\#\#\#}
\volnum{000}
\year{0000}
\pgrange{1--}
\setcounter{page}{1}
\lp{1}

\section{Introduction}
The noting by \cite{Curtis1918} that in the elliptical galaxy M87 ``a curious straight ray lies in a gap in the nebulosity ... , apparently connected with the nucleus by a thin line of matter'' marked the discovery of the first astrophysical jet, although its significance was not then recognized. The term `jet' was first used by \cite{Baade1954}   
who noted ``several strong condensations in the outer parts of the jet'' in M87 and also reported ``strong emission line of [O II] $\lambda$3727 \AA, which is shifted relative to the nuclear G-type spectrum by $-$295$\pm$100 km~sec$^{-1}$''. They suggested that this may be due to ejection from the nucleus. The jet in the quasar 3C273 was referred to as a ``faint wisp or jet'' while reporting the discovery of quasars \citep{Hazard1963,Schmidt1963}. These were the early beginnings. 

Although a number of jets in radio galaxies were mapped in the 1970s at radio frequencies \citep[e.g.][]{Northover1973,Turland1975,vanBreugel1977},
the ubiquity of radio jets was demonstrated from observations with the Very Large Array in the 1980s. This along with the subsequent detection of jets across the electromagnetic spectrum have helped develop a deeper understanding of astrophysical jets in active galactic nuclei or AGN \citep[e.g.][]{Harris2006,Blandford2019}.
These jets indicate the channels via which energy, momentum, mass and magnetic field are transported from the central supermassive black hole and its accretion disk to form the extended lobes of radio emission. As suggested by \citet{Bridle1984} we define a radio jet to be at least four times longer than its width. The jets range in size from sub-pc scales seen in the nuclear regions of active galaxies to hundreds of kpc for the largest radio sources.  

\begin{figure*}
    \centering
    \hbox{
    \includegraphics[width=9.25cm]{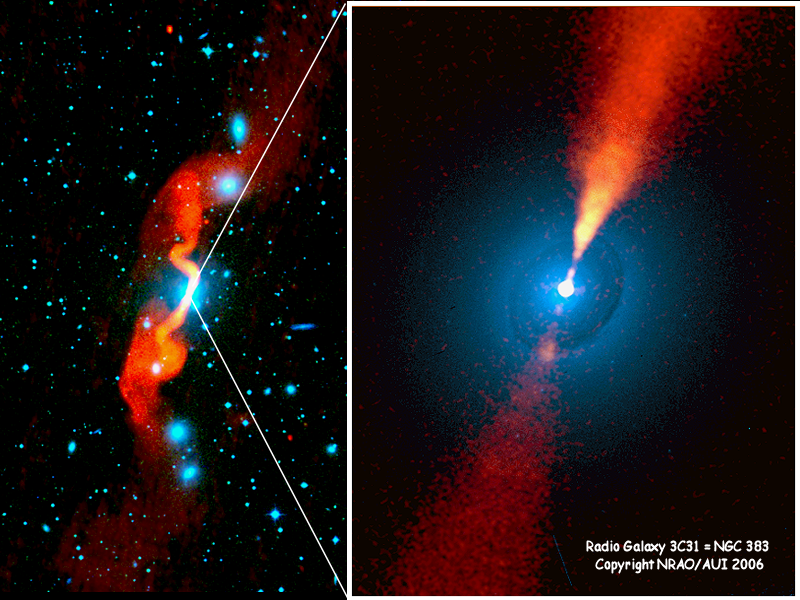}
    \includegraphics[width=7.3cm]{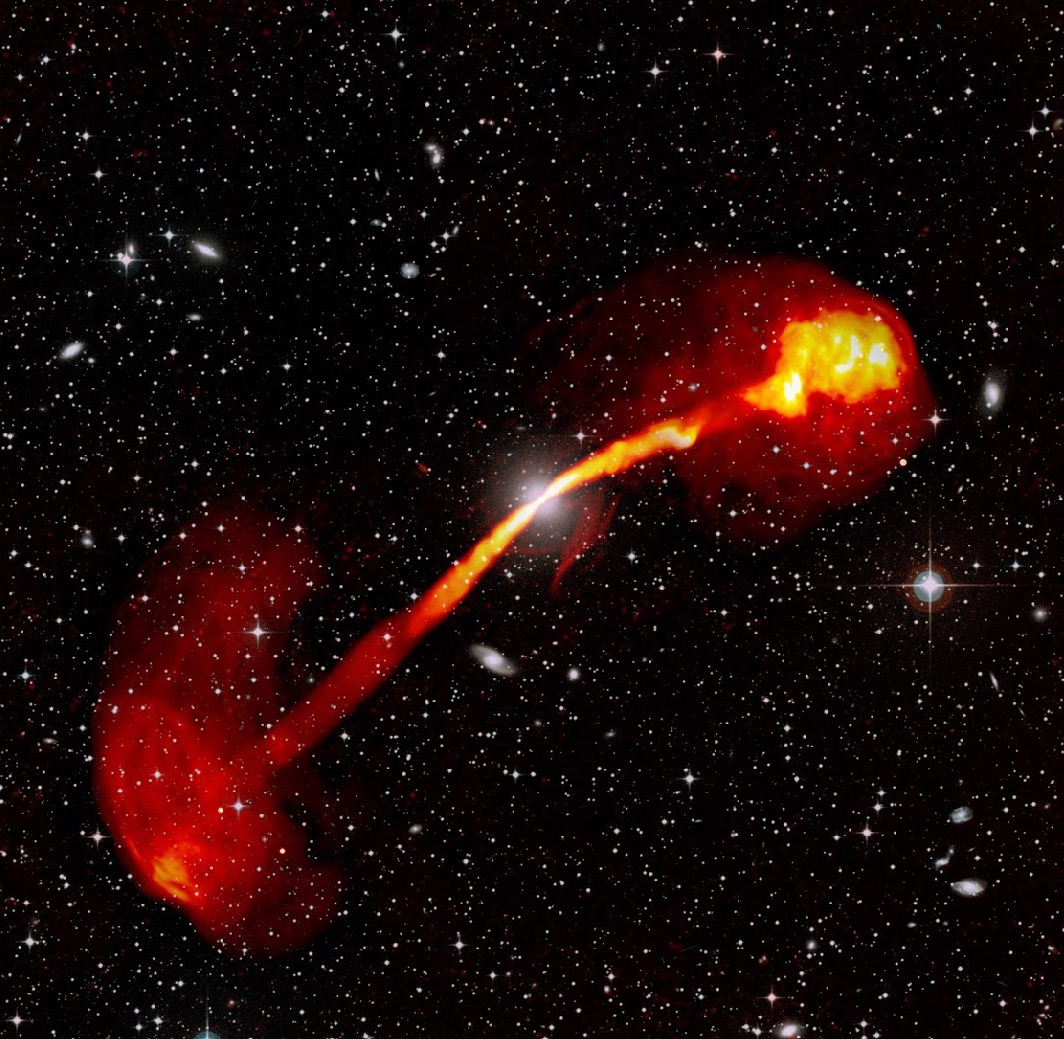}
    }
    \caption{Examples of Fanaroff-Riley Class I sources. Left: VLA radio image of the radio galaxy 3C31 shown in red and orange colours superposed on the Palomar Sky Survey optical image shown in blue. Middle: Higher-resolution VLA image of the inner jet superposed on the Hubble Space Telescope WFPC2 image. Right: MeerKAT radio image of the FR class I source IC4296 shown in orange and red hues superimposed on the SuperCOSMOS Sky Survey image in visible light. Credits for 3C31: NRAO, Alan Bridle; wide-field radio data: \cite{Laing2008}; HST/WFPC2 image from \cite{Martel1999}. Credits for IC4296: SARAO, SSS, S. Dagnello and W. Cotton (NRAO/AUI/NSF). Adapted from \cite{Condon2021}.
    }
    \label{f:3C31_IC4296}
\end{figure*}

\begin{figure}
	\centering
	\hbox{ 
		\includegraphics[width=8.5cm]{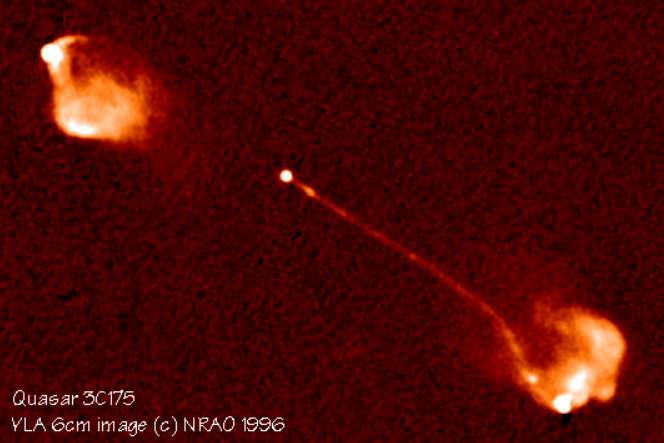}
	}
	\caption{Example of an FR class II source.  VLA radio image of the FR class II source 3C175 associated with a quasar. Credit: NRAO and Alan Bridle; adapted from \cite{Bridle1994}.
	}
	\label{f:3C175}
\end{figure}

The early developments in our understanding of radio jets were summarised by \cite{Bridle1984} in their seminal review. They noted that the jets in the lower-luminosity, edge-darkened sources without prominent
hot-spots at the outer edges (Fanaroff$-$Riley class I or FRI sources) tend to have two-sided radio jets although these may be one-sided
close to the parent optical object, while the higher-luminosity, edge-brightened sources with prominent hot-spots (FRII sources) tend to have one-sided jets (Figs.~\ref{f:3C31_IC4296} and \ref{f:3C175}). The traditional
dividing luminosity between these two classes identified by \cite{Fanaroff1974} is $\approx$10$^{26}$ W Hz$^{-1}$ at 150 MHz in a cosmology with H$_{o} = 70$ km s$^{-1}$ Mpc$^{-1}$, 
$\Omega_{\rm m}$ = 0.3 and $\Omega_\Lambda$ = 0.7. The magnetic field orientations also appear to be different for the jets in the two FR classes.
The jets in FRII sources which are either one-sided or highly asymmetric were found to exhibit a magnetic field predominantly parallel to the jet axes, while in the lower-luminosity FRI sources the magnetic field was either predominantly perpendicular to the jet axes or had a combination of perpendicular and parallel components \citep{Bridle1984}. 

Besides radio galaxies and quasars, radio jets have also been observed in Seyfert galaxies, an archetypal example being NGC4151 (\citealt{Williams2017}, and references therein), low-luminosity AGN (LLAGN) and also in star-forming H{\sc{ii}} galaxies \citep{Baldi2021}. The Seyfert, LLAGN and H{\sc{ii}} galaxies span the lower luminosity region of radio selected AGN. For example local radio luminosity function of AGN at 1.4 GHz extends down to a few times
10$^{20}$ W Hz$^{-1}$ \citep{Mauch2007}. The radio luminosity of Seyferts at 1.4 GHz lie in the range of a few times 10$^{20}$ to 10$^{24}$ W Hz$^{-1}$ \citep[e.g.][]{Ulvestad1989}. The luminosity distribution of the LOFAR Two-Metre Sky Survey, LoTSS-DR1, \citep{Shimwell2017,Shimwell2019} sources ranges from $\approx$10$^{21}$ to 10$^{29}$ W Hz$^{-1}$ at 150 MHz, the traditional dividing line between FRI and FRII sources being at 10$^{26}$ W Hz$^{-1}$ \citep{Mingo2019,Mingo2022}.

In addition to AGN, astrophysical jets have been found in a wide variety of cases, such as protostellar jets \citep{Bally2016}, pulsar wind nebulae \citep{Durant2013}, $\gamma-$ray bursts \citep{Gehrels2009}, stellar binary systems with a black hole companion such as SS433, and the micro-quasars which exhibit superluminal motion of the radio jets \citep{Mirabel1999}. Similar principles have been invoked in understanding these jets. 

In this article, we confine ourselves to jets in radio galaxies and quasars, with more emphasis on radio observations, summarising our current understanding and discussing future work. High-energy emission from jets and their spectral energy distributions are being discussed in an accompanying article in this issue \citep{Singh2022}, and have also been extensively covered in the reviews by \cite{Harris2006}, \cite{Worrall2009}, \cite{Blandford2019} and  \cite{Hardcastle2020}. From the vast body of literature 
we have been able to cite only a limited number of articles in this relatively short review.

\section{FR classes, HERGs and LERGs, radio jets}
The Fanaroff-Riley or FR classification of sources, with the two classes exhibiting different jet structures, was till recently based on studies of strong source samples, such as the 3CR and 2-Jy samples. From the available information at that time  \cite{Ledlow1996} found the dividing radio luminosity between the two classes to increase with optical luminosity of the host galaxy, although recent studies have shown the relationship to be more complex \citep[e.g.][]{Mingo2019,Mingo2022}. 

A number of reasons have been suggested for the observed dichotomy in the FR classes and their jets.
These can be broadly classified as (i) entrainment of
thermal material by the jets close to the nuclear region in FRI radio sources \citep[e.g.][]{Laing2007}; (ii) fundamental differences in the central engine such as the spin of the black hole and/or material forming the jet \citep[e.g.][]{Celotti1997}; and (iii) differences in the external environment and jet power which determine how rapidly jets may decollimate \citep[e.g.][]{GopalKrishna1996}. 
Studies of radio sources in different environments suggested that on average FRI sources tend to lie in higher density environments than FRII sources, indicating that jets may be affected by a denser surrounding medium \citep[e.g.][]{Wing2011,Gendre2013}. There is  observational evidence
that deceleration and decollimation of jets in FRI sources on small scales, with possible entrainment of material from the interstellar medium, may play an important role in the observed dichotomy \citep[e.g.][]{Bicknell1994,Laing2002a,Laing2002b,Laing2014,Mingo2019,Hardcastle2020}. 

Radio galaxies have also been traditionally classified based on their optical spectra since the early work by \cite{Hine1979}, into low-excitation radio galaxies (LERGs) and high-excitation radio galaxies (HERGs) \citep[e.g.][]{Hardcastle2007,Buttiglione2010,Best2012,Heckman2014,Tadhunter2016}.  In the low-excitation or jet-mode AGN, accretion is radiatively inefficient (RI) where the Eddington
ratio is less that 1\%, while in the high-excitation or radiative mode AGN, which is radiatively efficient (RE), the Eddington ratio
is greater than 1\%. In LERGs the nuclear region is
dominated by a ‘‘geometrically thick advection-dominated accretion flow’’ \citep{Narayan1995}, while in HERGs accretion is via the classical optically thin, geometrically thick accretion disk \citep{Shakura1973}. These aspects have been summarised by \cite{Heckman2014} in their review.
 Traditionally the LERGs  have been found to have an FRI-type structure although there are a significant number of FRII LERGs, 
while HERGS are predominantly of FRII type \citep{Best2012,Heckman2014,Tadhunter2016}. Recent studies from deep radio surveys with LOFAR have significantly altered our understanding of the relationship between FR class, accretion mode and host galaxy properties \citep{Mingo2019,Mingo2022}.

\begin{figure}
	\centering
	\vbox{ 
		\includegraphics[width=8.5cm]{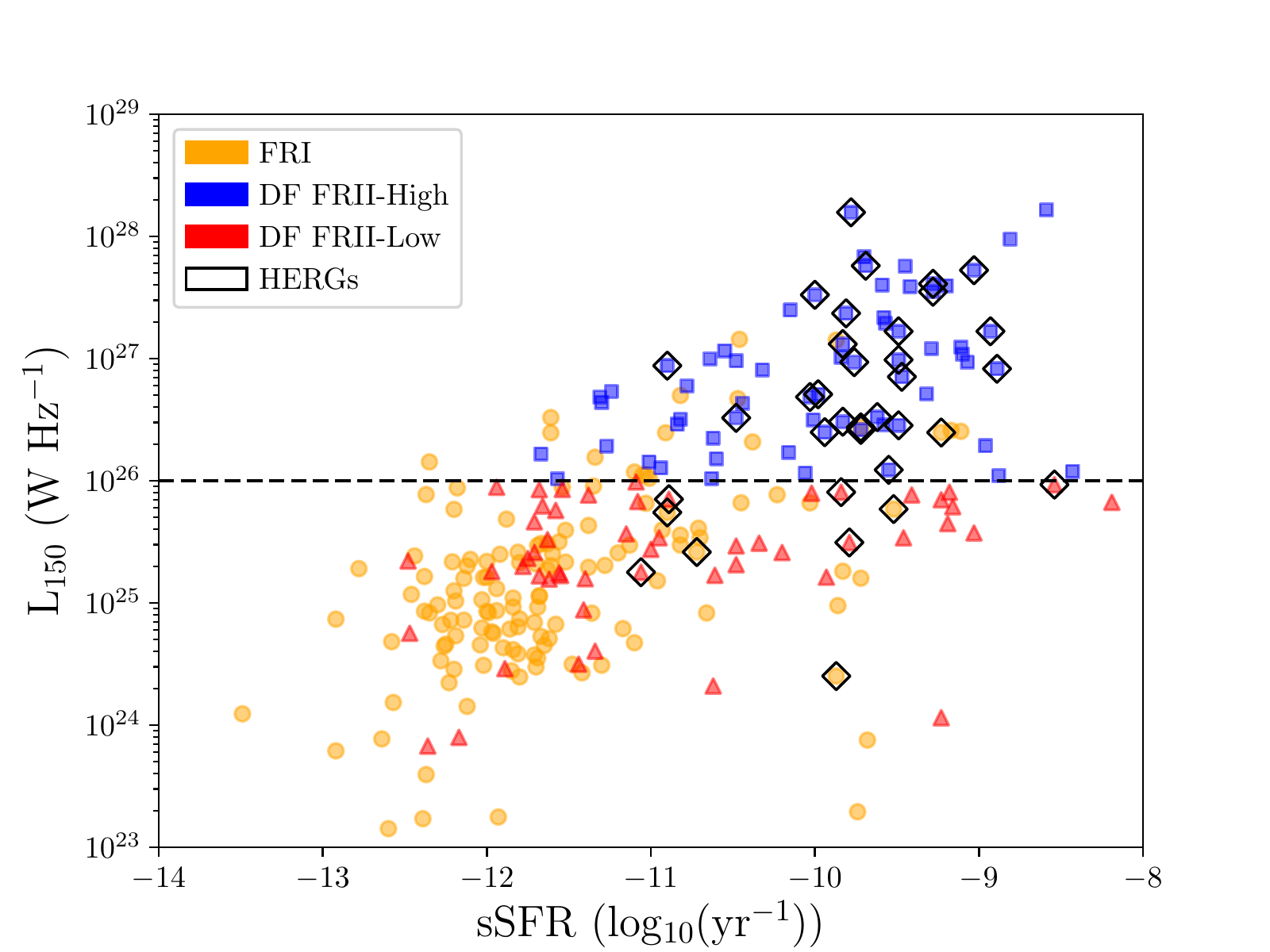}
		\includegraphics[width=8.5cm]{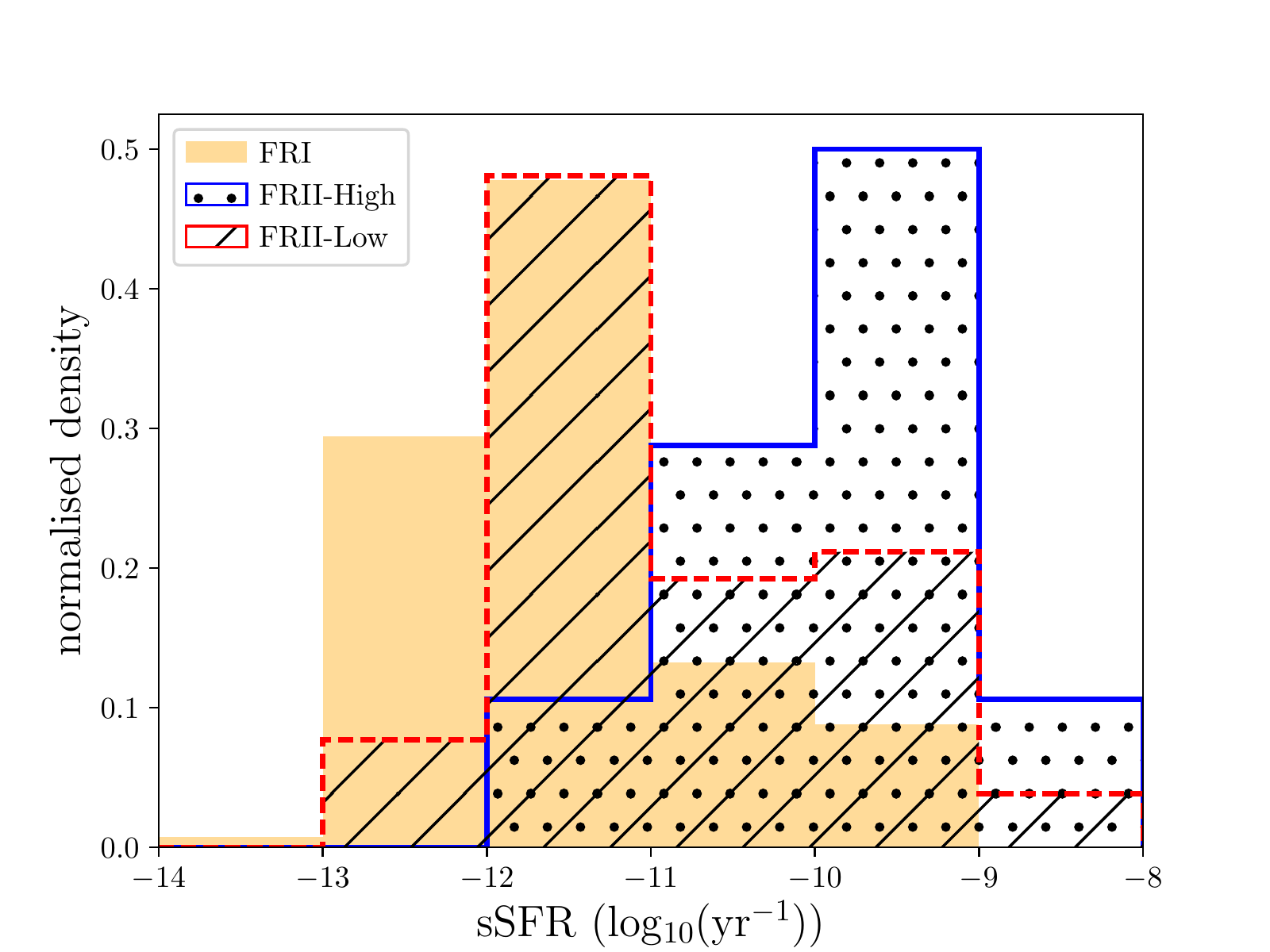}
		\includegraphics[width=8.5cm]{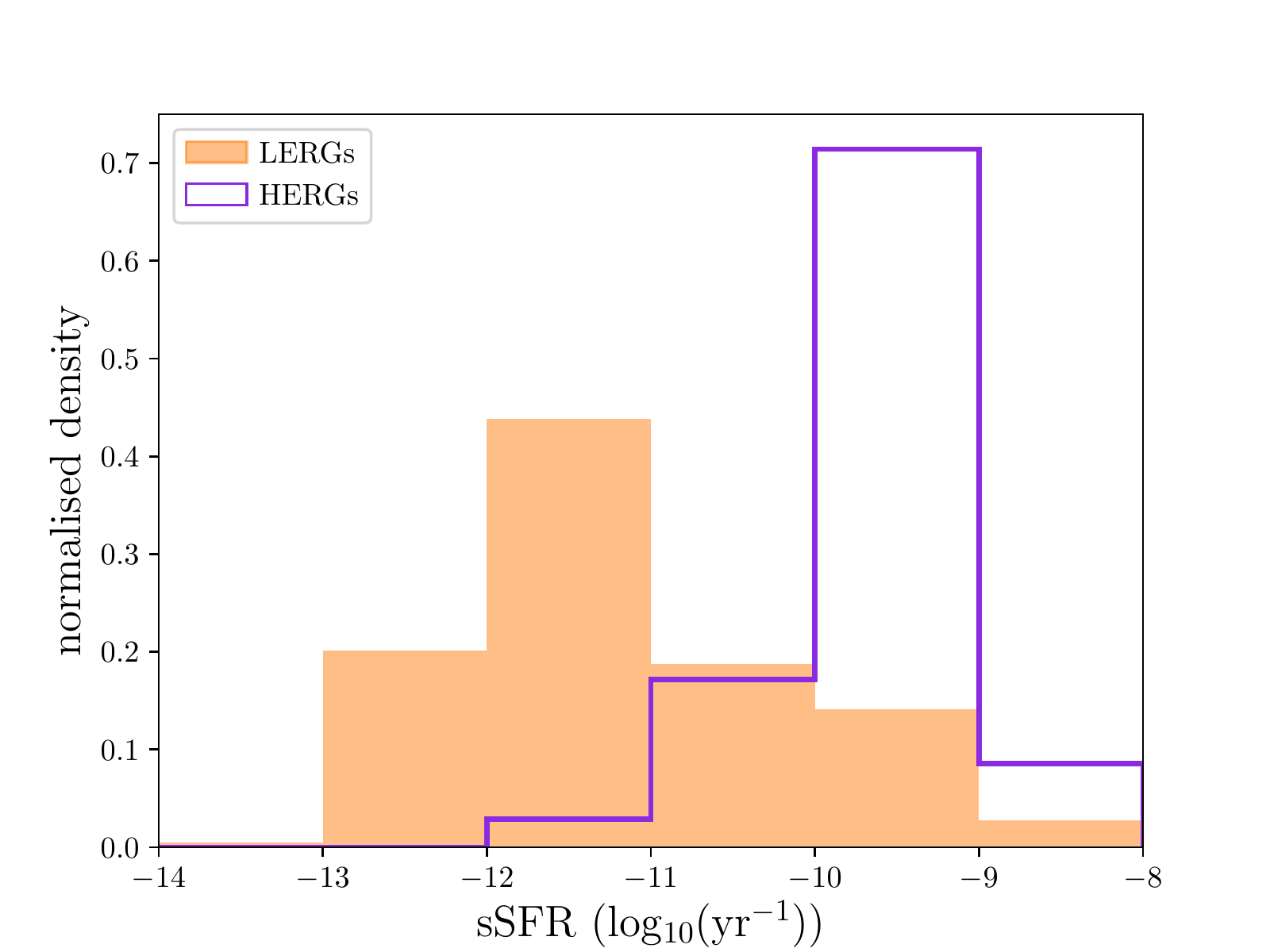}
	}
	\caption{Upper panel: Specific star formation rate (sSFR) vs radio luminosity at 150 MHz for FRIs and low- and high-luminosity FRIIs as indicated in the figure. The HERGs are indicated. Middle panel: Distributions of sSFR for the FRIs and low- and high-luminosity FRIIs. Lower panel: Distributions of sSFR for HERGs and LERGs. Figures are from \citet{Mingo2022}.}
	\label{f:mingo2022}
\end{figure} 

We briefly summarise a few of the significant results of \citet{Mingo2022} which have a significant bearing in our understanding of jet formation, accretion mode and large-scale radio structure (Fig.~\ref{f:mingo2022}).
Dividing the FRIIs into FRII-high and FRII-low in the traditional dividing line of L(150 MHz) 
= 10$^{26}$ W Hz$^{-1}$ for the FRI and FRII classes, they find that $\sim$65 per cent of the FRII-high sample are LERGs, contrary to earlier studies. There appears to be no significant difference in the large-scale radio structure on 100-kpc scale between FRII LERGs and HERGs, suggesting that FRII ``classification is not primarily controlled by the central engine''. As in earlier studies, they find a significant population of FRIIs below the dividing luminosity, suggesting that FR classification is not determined by jet power alone. FRII sources appear across all luminosities and both accretion modes. Low-luminosity FRIIs and FRIs are overwhelmingly LERGs, so that RE accretion is rare at these luminosities. By comparing low-luminosity FRIIs and FRIs of similar luminosity, they show that the probability of a low-power jet becoming either an FRI or FRII jet depends on the stellar mass of the host galaxy. This would be consistent with the ideas of the environment playing an important role in the formation of FRI jets. HERGs across all luminosities and morphologies tend to have high specific star formation rates, suggesting a close link with availability of fuel. Radio morphology and jets, accretion mode and host galaxy properties appear related but in more complex ways than simple one-to-one relationships \citep{Mingo2022}. 

These results raise a number of interesting questions. Traditionally FRI and FRII sources divided by luminosity have shown evidence of different evolutionary properties \citep[e.g.][]{Wall1980a,Wall1980b}. If the luminosity-FR class division is blurred, is the primary dependence of evolution on FR class or luminosity or the LERG/HERG classification? The radio jet structures of FRI and FRII sources are also different. Although low-luminosity FRIIs are also expected to have reasonably well-collimated jets as the hot-spots are visible, are their jet structures and field orientations similar to those of high-luminosity FRIIs? The internal composition of the relativistic plasma in the lobes of FRI and FRII radio sources appear to be different \citep[e.g.][]{Croston2018}. How does this extend to low-luminosity FRIIs? Are the jets in low-luminosity FRIIs more susceptible to instabilities and entrainment than high-luminosity FRIIs? As these low-luminosity sources are imaged with greater sensitivity and resolution to clarify their jet structures, it would be interesting to pursue some of these questions.

The evolution of LERGs and HERGs and their impact on galaxy evolution also need to be better understood. Recently, \citet{Kondapally2022} have examined this aspect for LERGs by splitting the sample into quiescent and star-forming galaxies. They find that the quiescent LERGs dominate the radio luminosity function at z$<$1 and are consistent with accretion occurring from cooling of hot gas halos. The star-forming radio luminosity function increases with redshift, dominating the space densities by z$\sim$1. They suggest that accretion in these cases is possibly due to cold gas present in these star-forming galaxies.

\section{The FR0 sources}
Combining sensitive radio surveys with optical surveys such as the Sloan Digital Sky Survey or SDSS \citep[e.g.][]{Best2012} has revealed a population of radio sources whose core luminosity is similar to that of FRI radio sources, but the extended emission is weaker by a factor of $\sim$100 \citep[e.g.][]{Baldi2009,Baldi2015,Sadler2014,Sadler2016,Cheng2018}.  These sources termed as FR0s can be found in both high- and low-frequency surveys. For example in the 
AT20G-6dFGS sample,  $\sim$68 per cent of the sources fall into the FR0 category \citep[e.g.][]{Sadler2014,Sadler2016}. Similarly for a complete sample of sources chosen from the Cambridge 10C survey at 15.7 GHz, again $\sim$68 per cent have been classified as FR0s \citep{Whittam2016}.  At low radio frequencies $\sim$70 per cent of the sources in LoTSS appear unresolved \citep{Shimwell2017,Shimwell2019}. In observations of deep fields at low frequencies such as of ELAIS-N1 the majority of sources are unresolved with an angular resolution of a few arcsec and have steep radio spectra \citep[e.g.][]{Sirothia2009b,Ishwara-Chandra2020}, 
all showing that FR0s are quite common among radio AGN at low luminosities.

A catalog of 108 FR0 sources (FR0CAT) with redshifts less than 0.05 and projected linear size $<$5 kpc was compiled by \cite{Baldi2018}. The host galaxies of the FR0s are massive luminous early type
galaxies $(-21 < M_r < -23)$ with mid-IR colours consistent with those of elliptical galaxies, and black hole masses of $\sim10^{7.5}$ - $10^9 M_\odot$, which are less than those in FRI radio galaxies. The most striking difference with FRIs is that the radio luminosity is lower than 3CR sources by $\sim$100 even for sources of similar [O{\sc{iii}}] luminosity \citep{Baldi2018}.  There are also indications that the galaxy density of FR0s may be lower than that of FRIs by a factor of $\sim$2 \citep{Capetti2020}. 

In this review we focus on the launching of jets in FR0s. Although FR0s are likely to be a mixed bag of objects, high-resolution observations often reveal evidence of jet-like features \citep[e.g.][]{Cheng2018,Baldi2019,Baldi2021}.  Although these features may not always be consistent with the classical definition of \cite{Bridle1984} of being 4 times longer than the width, they do provide evidence of collimated ejection of relativistic plasma from an AGN. A search for extended emission in FR0s observed with LOFAR show that about 20 per cent show evidence of bipolar emission on opposite sides \citep{Capetti2020}.

 \cite{Cheng2018} observed 14 FR0s with VLBI techniques and found 4 of the sources to have Doppler boosting factors ranging from 1.7 to 6, and two with multi-epoch observations to have proper motions between 0.23 and 0.49c. \citet{Baldi2021b} report high-resolution observations of 15 FR0s with eMERLIN, EVN and JVLA and find that most show evidence of jet-like structures. \citet{Baldi2021b} also report a linear correlation between the radio core luminosity and  [O{\sc{iii}}] line luminosity for a sample of low-luminosity active nuclei consisting of both FR0s and FRIs, suggesting similar disk-jet coupling for these sources. The high-resolution studies are consistent with FR0s having mildly relativistic jets. 

The similarity of host galaxies of FR0s and FRIs does not suggest that the jets in FR0s are confined to small dimensions by a dense medium. Possible reasons suggested for understanding the inability to launch large-scale jets as in FRI sources include low black hole mass \citep{Miraghaei2017} and/or black hole spin (\citealt{Baldi2021b}, and references therein).  
Theoretical studies indicate jet power could depend strongly on the black hole spin and may also provide a viable explanation for the radio loud - radio quiet dichotomy \citep[e.g.][]{Tchekhovskoy2010}. \citet{Maraschi2012}
suggest that above a spin threshold, black hole spin and accretion rate could lead to a grand unification of AGN. Observationally, the detection of maximally rotating black holes in the low-luminosity Seyfert galaxies (e.g. Table 2 in \citealt{Brenneman2011}) suggests that spin alone may not be the determining factor for inability to launch high-luminosity radio jets. It is 
possibly due to a combination of black hole mass, spin and accretion rate, which requires more observational and theoretical work to clarify.

\section{Hybrid morphology sources and radio jets}
While detailed studies of jets in FRI and FRII sources are discussed later, here we discuss briefly the nature of sources which appear to have a hybrid morphology. These are sources where one side appears to have an FRI structure while the opposite side exhibits an FRII structure. 

\begin{figure}
    \centering
    \includegraphics[width=8.5cm]{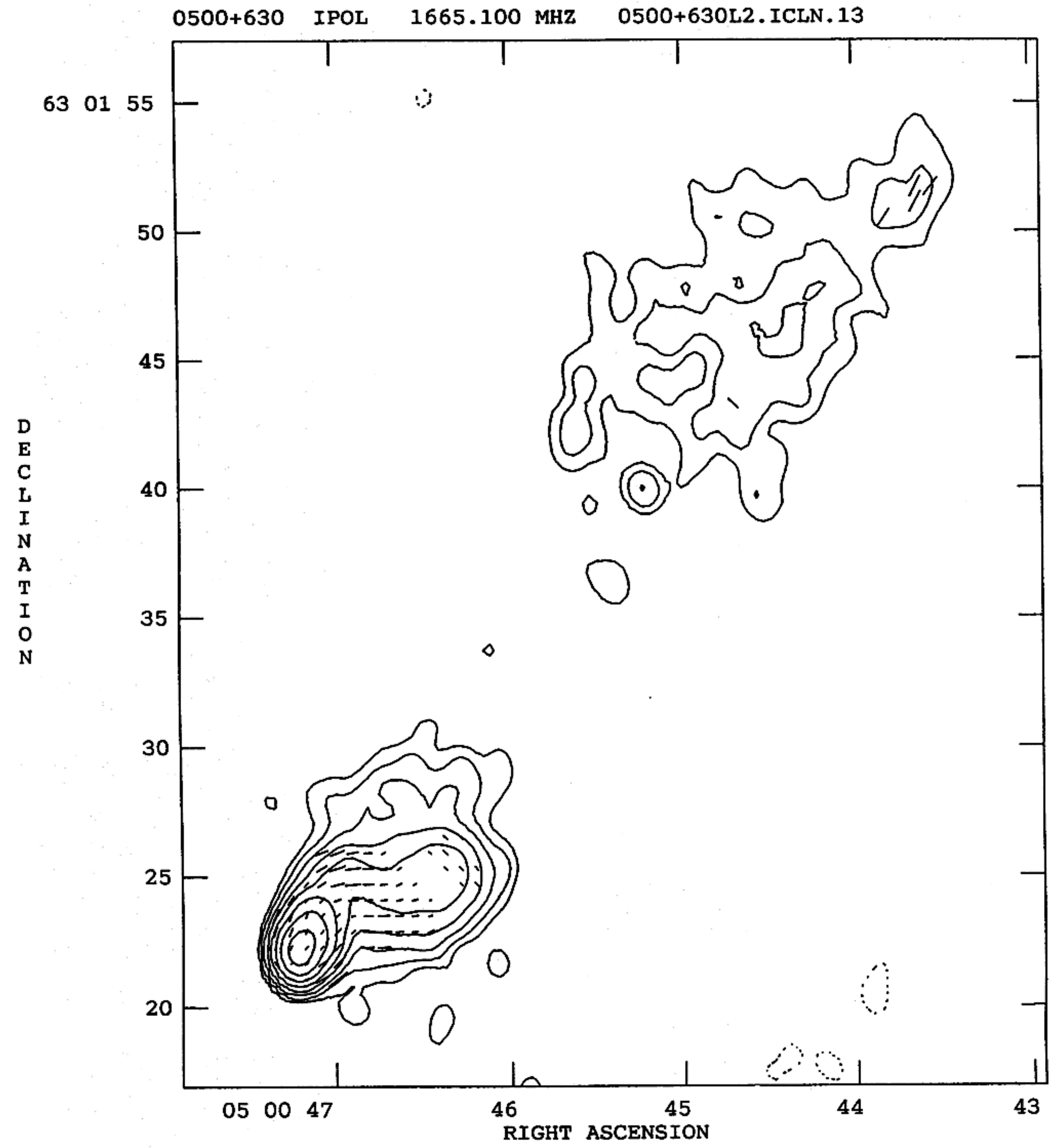}
    \caption{A highly asymmetric radio galaxy with an apparent hybrid FRI-FRII morphology \citep{Saikia1996}.}
    \label{f:0500+630}
\end{figure}

One of the very early examples of these sources 
is B0500+630 (Fig.~\ref{f:0500+630}) where the authors noted that ``the source appears to have a composite structure, with one side being typical of Fanaroff-Riley class II sources, while the diffuse lobe is similar to those seen in Fanaroff-Riley class I sources'' \citep{Saikia1996}. However, although no hotspot is visible on the apparently FRI side, there is also no evidence of a jet, symptomatic of jets in FRI sources, connecting the core to the diffuse lobe. Being associated with a galaxy B0500+630 is likely to be inclined at $>$45$^\circ$ to the line of sight. With a peak brightness ratio of the oppositely-directed hotspots of $\sim$100, the source appears to be intrinsically asymmetric. \cite{GopalKrishna2000} compiled a sample of 5 such sources and suggested that this supports the scenario that the FR dichotomy is due to jet interaction with the external environment rather than due to differences in the central engine, such as black hole spin, or differences in jet composition. 

From existing surveys of sources further examples of such sources, termed HyMoRS, were reported by a number of authors \citep[e.g.][]{Gawronski2006,Banfield2015,Kapinska2017}. \citet{Ceglowski2013}  observed a sample of 5 HyMoRS using the Very Large Baseline Array (VLBA) and found core-jet structures in two of them, one pointing towards an FRII-like lobe and the other towards an FRI-like one, and two probable weak jets. They suggested that HyMoRS are possible ``FRIIs evolving in a heterogeneous environment''. 
More recently, \citet{Harwood2020} made a detailed study of a small sample of HyMoRS examining their spectral index distributions and the injection spectral indices. They concluded that these ``objects are most likely the result of
orientation and are intrinsically FRII radio galaxies''.

High-resolution sensitive observations of candidate HyMoRS to examine both their spectra and structure in both total intensity and polarization would be helpful to clarify whether there are genuine HyMoRS. 
Absence of hotspots alone on one side may not be adequate to classify a radio galaxy as a HyMoRS. The total-intensity and polarization structure of the radio jets, besides spectral index information, could provide valuable clues towards identifying genuine HyMoRS. 
There could be intrinsic asymmetries in well-collimated jets in high-luminosity FRII sources with significantly weaker hotspots on one side. For example, the hotspots in the high-luminosity quasars 3C9, 3C280.1 and B1857+566 have very weak hotspots on the side facing the jets, which are possibly approaching us within about 45$^\circ$ to the line of sight \citep[cf.][]{Swarup1982,Saikia1983}.  

\section{Nuclear or VLBI-scale jets}
As summarized by \citet{Blandford2019}, nuclear or VLBI-scale jets tend to be one-sided with a flat-spectrum nuclear core at one end, and with components often appearing to move along the jet with superluminal velocities. Superluminal motion is common in core-dominated radio sources which are inclined at small angles to the line of sight, with apparent velocities ranging from $\sim$0.03c to 50c. Surveys of jets on VLBI scales in the last couple of decades include the 
Very Long Baseline Array Calibrator Survey \citep{Beasley2002}, Australian Long Baseline Array survey of southern sources \citep{Petrov2019}, Monitoring Of Jets in Active galactic nuclei with VLBA Experiments or MOJAVE \citep{Lister2016,Lister2018} and monitoring a sample of $\gamma$-ray blazars \citep{Jorstad2017}. The Astrogeo Project contains the observations of about 12000 AGN observed by VLBI techniques \citep{Petrov2022}.

The polarization properties of nuclear or parsec-scale jets have also been studied using VLBI techniques. One of the extensive studies based on observations of 484 sources over the time interval 1996-2016 was reported by \citet{Pushkarev2017}. They report a significant increase in the degree of linear polarization with distance
from the radio core along the jet for quasars, BL Lac objects and galaxies, and also an increase towards the edges of the jets. The increase with distance could be due to more ordered fields further down the jet, while the increase towards the edges is possibly due to greater depolarization closer to the jet axes. The cores and jets of BL Lac objects tend to be more polarized than quasars. Also the E-vector position angles (EVPA) of the cores tend to be more stable in BL Lacs and the EVPAs in both the cores and jets appear better aligned with the jet axes. This suggests compression of the magnetic field due to shocks with the B-field being perpendicular to the jet direction. \citet{Pushkarev2017} found no such trend for the jets in radio galaxies and quasars.

One of the most extensive studies of rotation measure (RM) estimates in the jets has come from the MOJAVE group who reported observations of 191 extragalactic jets \citep{Hovatta2012}. They find the quasars to have on average larger RM values than BL Lacs, and the cores to have higher values than the jet components. There is a significant negative correlation between the jet RM and deprojected distance from the core. They find significant transverse RM gradients in 4 sources, with the RM in the quasar 3C273 changing sign from positive to negative along the transverse cut. This result was confirmed by \citet{Wardle2018} who estimate a current of $10^{17}$-$10^{18}$ A flowing down the jet. The RM variations transverse to the jet indicates a toroidal field, although the field is largely along the axis of the jet. Wardle also find the RM distribution to be variable on time scales of months to years and suggest that this is due to motion of superluminal components behind a turbulent Faraday screen around the jet. ALMA observations at 1 mm on a scale of about 2 kpc suggest a sheath surrounding a conically expanding jet \citep{Hovatta2019}. 
Gradients in RM transverse to the jet axes which could be due to helical or toroidal magnetic fields have been reported for a number of other AGN as well \citep[e.g.][]{Gabuzda2015,Gabuzda2021,Kharb2009}.  These could play a significant role in the collimation of the jets.

\begin{figure}
    \centering
    \includegraphics[width=8.5cm]{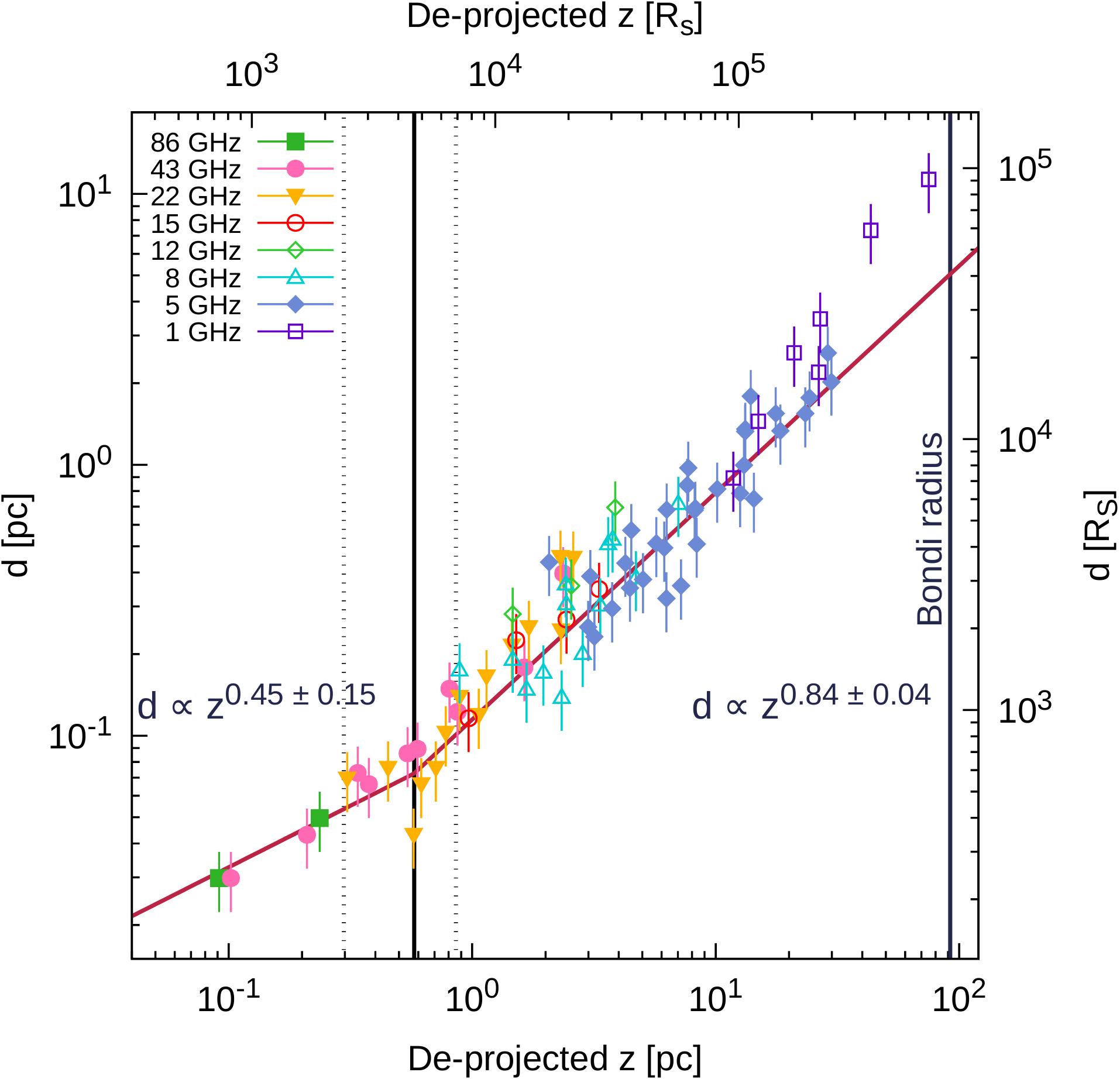}
    \caption{The jet collimation profile in NGC315 illustrating the transition from a parabolical  shape to a conical one on sub-parsec scale \citep{Boccardi2021}. Reproduced with permission \copyright ESO.}
    \label{f:NGC315}
\end{figure}

Observational studies of the collimation of jets are important to understand the physics of jets, including their formation, propagation and acceleration processes.
High-angular resolution observations are required to determine the jet profiles and their variation with distance from the radio core. The variation of the apparent jet width $w$ with distance from the core $r$ is usually fitted with the function of the form $w \propto r^k$, where $k \approx 0.5$ for a quasi-parabolic jet and 1 for a conical jet. One of the early evidences for transition from a parabolic to a cylindrical jet shape was in the radio galaxy M87, the transition occurring near the feature HST-1 at a projected distance of $\approx$70 pc, corresponding to 
$\approx 10^5$R$_s$, where R$_s$ the Schwarzschild is given by $2GM/c^2$ \citep{Asada2012}.  \citet{Pushkarev2017}  found most resolved jets to have an approximately conical shape. 
Observational evidence suggests a jet strucure with a fast spine and a slower outer layer. 
\citet{Hervet2017} attempt to link different types of AGN with specific stratified jet characteristics based on VLBI observations of a large sample of AGN jets.
A number of other authors have attempted to study the jet profiles in the innermost jet regions (e.g. \citealt{Kovalev2020}, and references therein).  \citet{Kovalev2020} find the transition from parabolic to cylindrical shapes to be quite common in AGN jets. The transition occurs at gravitational radii, $r_g = GM/c^2 \approx 10^5 - 10^6$, which roughly corresponds to the Bondi radius $r_B = 2GM/c_s^2$ where $c_s$ is the sound speed. They suggest that the transition occurs where the bulk plasma kinetic energy equals the Poynting energy flux, with Bondi accretion determining the pressure of the ambient medium. Detection of features in the jets possibly due to shocks at the transition region where the jets become plasma dominated appears to support this scenario \citep{Kovalev2020}.

Although change from a parabolic to conical jet collimation profile around the Bondi radius appears fairly common, there are also examples of deviation from this picture. An interesting example is the low-ionisation nuclear emission line region (LINER) galaxy NGC1052 with twin-jets. \citet{Baczko2022} 
find that both jets are conical downstream of a break in the  jet collimation profile at $10^4$R$_s$.
However upstream of the jets, the jet collimation profile is neither cylindrical nor parabolic for the approaching jet and close to cylindrical for the receding one. While more observational work is required, evidence of differences in collimation on opposite sides in the nuclear jets will also have implications in interpreting asymmetries in the large-scale structure.  

We highlight briefly a few significant results from recent studies of radio jets on parsec or sub-parsec scales in different sources which have been reported since the review by \citet{Blandford2019}.

\subsection{Giant radio galaxy NGC315, J0057+3021}
NGC315 is a giant radio galaxy with a black hole mass of $\approx 1.3\times 10^9 M_\odot$, whose sub-parsec scale structure has been studied recently by \citet{Boccardi2021} and \citet{Park2021}.  \citet{Boccardi2021} have observed it with higher resolution extending to 86.2 GHz using the Global Millimeter-VLBI Array (GMVA). Both groups find a transition from a parabolic to a conical shape, although \citet{Boccardi2021}  find it to occur closer to the central engine at a distance of 0.58$\pm$0.28 pc or $\sim 5\times10^3$ Schwarzschild radii (Fig.~\ref{f:NGC315}). This is much smaller than the Bondi radius which has been estimated to be 92 pc from x-ray observations. The transition appears to occur at sub-pc scales, after which it remains conical to kpc scales. They note a similar behaviour in other low-luminosity AGN (e.g. NGC4261, Cen A) and suggest that the initial confinement of the jet may be due to a thick disk extending $\sim 10^3$-$10^4 R_s$.

\begin{figure}
    \centering
    \includegraphics[width=8.5cm]{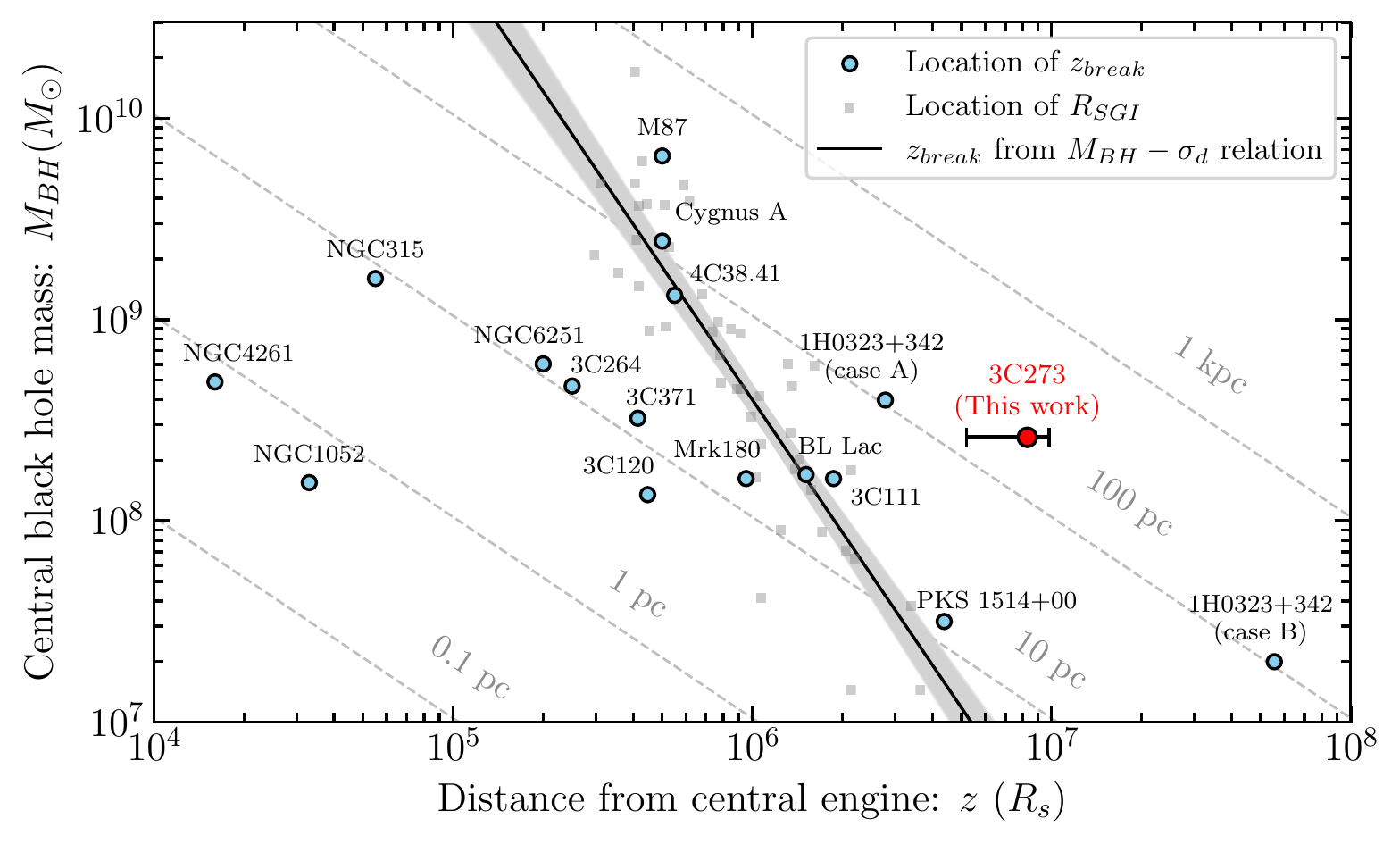}
    \caption{The relation between the central black hole mass and the de-projected distance of the jet collimation break from it \citep{Okino2021}.
    }
    \label{f:Okino_3C273}
\end{figure}

\begin{figure*}[th]
    \centering
    \includegraphics[width=10cm]{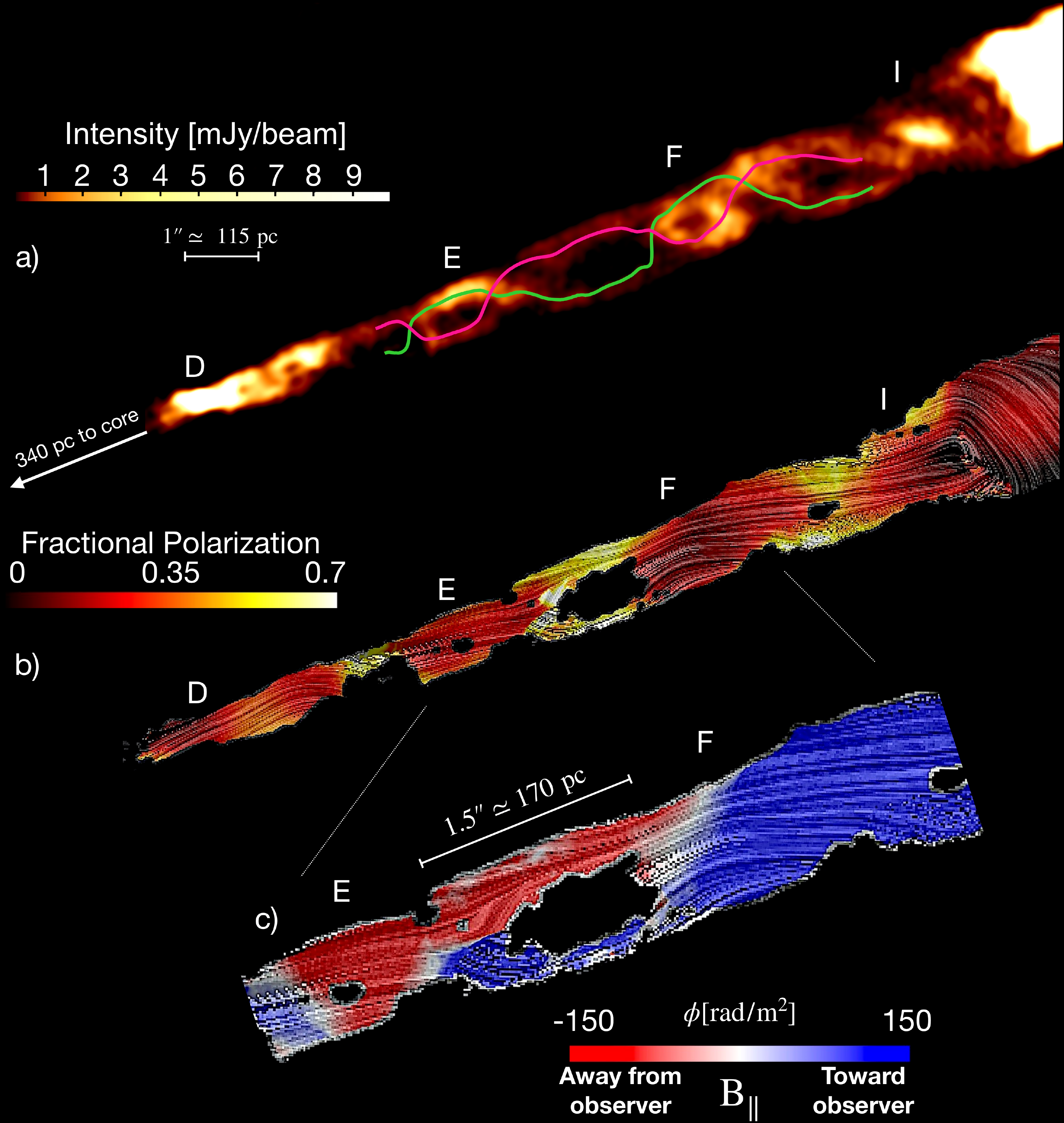}
    \caption{A polarization study of the conical jet of M87 reveal a helical magnetic field configuration \citep{Pasetto2021}. The upper panel shows the 
    double-helix structure between knots D and I. The middle image shows the magnetic field lines largely follow the double-helix structure. The bottom figure which plots the Faraday depth shows that the magnetic field has opposite directions where they are clearly able to separate the emission from both edges. These suggest a helical configuration for the M87 jet (see \citealt{Pasetto2021} for more details). 
    }
    \label{f:M87_Pasetto}
\end{figure*}

\begin{figure}
    \centering
    \includegraphics[width=8.5cm]{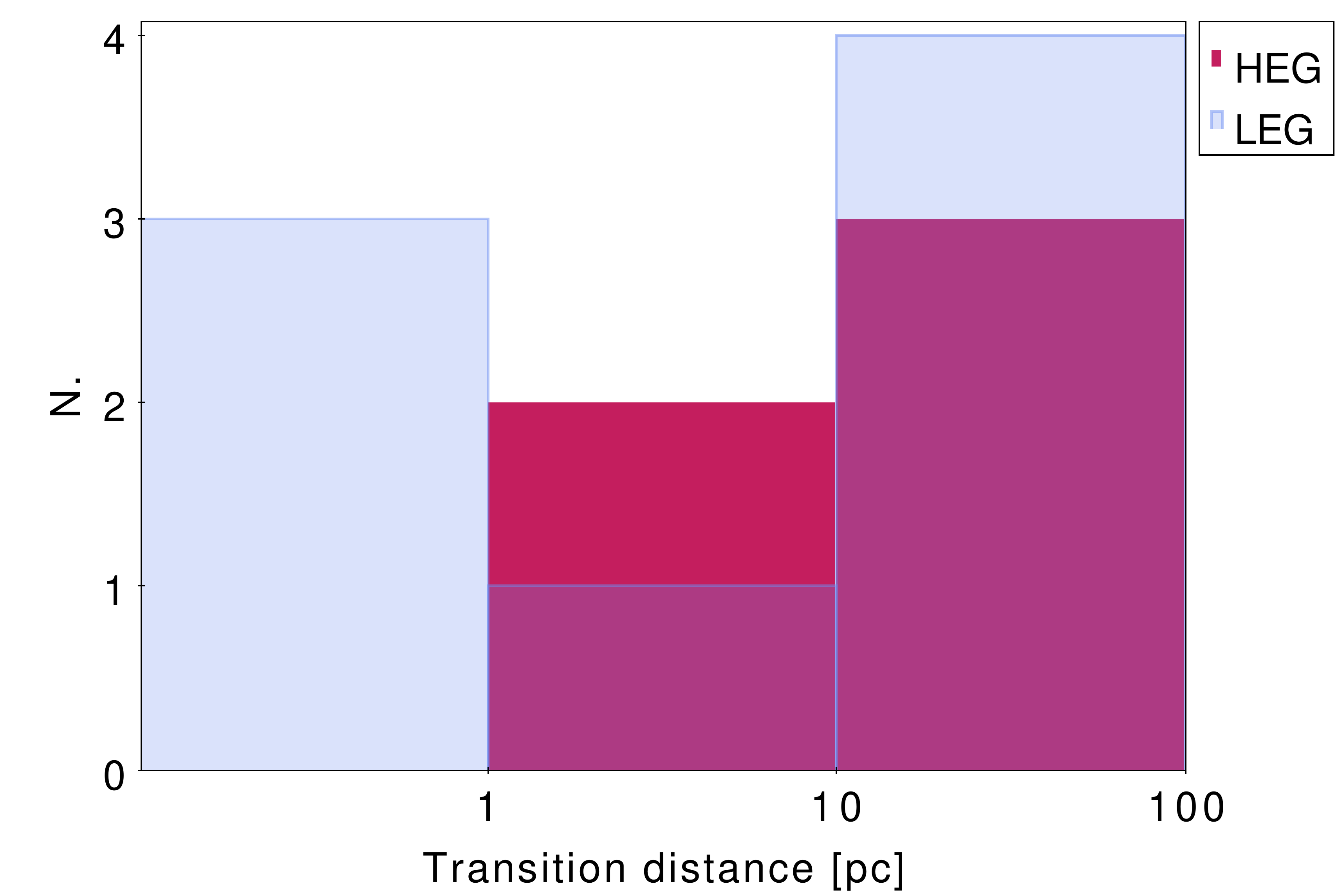}
    \caption{Distribution of the de-projected transition distances in  LERGs and HERGs where the distance is expressed
    in units of their Schwarzschild radii \citep{Boccardi2021}. Reproduced with permission \copyright ESO.}
    \label{f:Boccardi_heg_leg}
\end{figure}

\subsection{Quasar 3C273, J1229+0203}
Is the transition in jet collimation seen in low-luminosity AGN also present in higher-luminosity sources with more powerful jets? The archetypal quasar 3C273, whose black hole mass and viewing angle have been estimated to be ($2.6\pm1.1)\times 10^8 M_\odot$ and 12$^\circ$ respectively by the \citet{GravityCollaboration2018}, was observed by \citet{Okino2021} with an angular resolution of 60 $\mu$as at 86 GHz. They resolve the innermost jet on scales of $10^5$ Schwarzscild radii, and find a similar behaviour to that of the lower-luminosity AGN. Here too the inner jet collimates parabolically while the outer jet expands conically. The jet collimation break is seen at $\sim 8 \times 10^6 R_s$, the Schwarzschild radius.

\citet{Okino2021} compare the results for 3C273 with other jets in AGN by exploring the relation between the deprojected distance of the collimation break vs the black hole mass. They find that the collimation break occurs over a wide range, from $\sim 10^4 R_s$ to $\sim 10^8 R_s$ (Fig.~\ref{f:Okino_3C273}). They suggest that the transition region is determined not merely by the sphere of gravitational influence of the black hole, but also diverse environmental factors such as the torus, disk, disk wind, and a hot gas cocoon around the jet. 

\subsection{FRI radio galaxy Centaurus A, NGC5128, J1325-4301}
\citet{Janssen2021} have presented the image of the nuclear jets in Cen A with an angular resolution of 25 $\mu$as, probing the structure of the jets at about 200 gravitational radii from the $5.5\times 10^7 M_\odot$ black hole. These observations reveal a highly collimated, asymmetrically edge-brightened radio jet as well as a weaker counter jet. There appears to be no radio emission from the spine of the jet, the sheath to spine intensity ratio being $>$5. They find the jet to have a wide initial opening angle of $>40^\circ$ and the width to vary with distance with $k = 0.33$. They suggest that this either indicates strong magnetic collimation or external ambient pressure and density decreasing as $\propto r^{-1.3}$ and $\propto r^{-0.3}$
respectively. The similarity of the spine-sheath structure and a large initial opening angle seen in other nearby galaxies, M87 \citep{Kim2018}, Mkn501 \citep{Piner2009} and 3C84 \citep{Giovannini2018} suggests that this may be a common feature in low-luminosity AGN. These very high-resolution observations provide an opportunity of comparing the observations with general relativistic magnetohydrodynamics (GRMHD) simulations \citep[e.g.][]{Chatterjee2019}. The sheath is possibly the region of interaction between the fast spine and the accretion powered outflow \citep{Janssen2021}.

\subsection{M87, OJ287 and 3C279}
In addition to the ones discussed above, there have been interesting results on jets obtained for a number of well-known AGN. From high-fidelity images of M87 with a resolution of 10 pc, \citet{Pasetto2021} find ``a double-helix morphology of the jet material between $\sim$300 pc and $\sim$1 kpc''. They suggest a helical magnetic field which is sustained on these scales by Kelvin-Helmholtz instabilities (Fig.~\ref{f:M87_Pasetto}). 

In the context of M87, it is important to note that since the publication of the total-intensity images around the supermassive black hole by the Event Horizon Telescope (EHT), linear polarization images have been reported recently \citep{Akiyama2021a,Akiyama2021b}. The high angular resolution of $\sim$20 $\mu$as, $\approx$2.5 R$_s$, enabled a study of the polarization properties, magnetic fields and plasma properties in the vicinity of the event horizon. Only a part of the ring appears polarized with the degree of polarization rising to $\sim$15 per cent in the south-western part. The low polarization is possibly due to unresolved structures within the EHT beam which they attribute to Faraday rotation within the emission region. The net linear polarization pattern is azimuthal which may be due to organized poloidal magnetic fields. The EHT Collaboration estimate the density n$_{e} \sim 10^{4-7}$ cm$^{-3}$, magnetic field strength B$\sim$ 1-30 G, and electron temperature T$_{e} \sim (1-12)\times10^{10}$ K. They also find that the consistent GRMHD models are of magnetically arrested accretion disks, and estimate a mass accretion rate onto the black hole of $(3-20) \times 10^{-4}$ M$_\odot$ yr$^{-1}$.
 
Polarimetric space VLBI observations of the well-known blazar OJ287 enabled imaging the inner jet with an angular resolution of 50 $\mu$as \citep{Gomez2022}. They find the innermost jet to be dominated by a toroidal magnetic field, and suggest that the VLBI core is threaded by a helical magnetic field. Another archetypal blazar 3C279 was observed in total intensity at mm wavelengths with an angular resolution of 20 $\mu$as \citep{Kim2020}. These observations show non-radial motion of inner jet components at apparent speeds of $\sim$15c and $\sim$20c.

\subsection{Collimation in LERGs, HERGs, FRI and FRII sources}
As mentioned earlier the collimation break occurs over a wide range, from $\sim 10^4 R_s$ to $\sim 10^8 R_s$ \citep[e.g.][]{Okino2021}. In NGC315 jet collimation is complete within a parsec, in M87 the jet is anchored in the vicinity of the ergosphere (\citealt{Kim2018} and references therein), while in Cygnus A, \citet{Boccardi2016} find a minimum jet width of $\sim230 R_s$ and suggest that the jet may be launched from a larger distance. The jet collimation break in 3C273 is also at a large distance of $\sim 8 \times 10^6 R_s$ \citep{Okino2021}.

Cygnus A is an archetypal FRII source and its jet power is larger than that of M87, an FRI source, by about 3 orders of magnitude. Do the nuclear jets suggest a difference in collimation between FRI and FRII sources, and between HERGs and LERGs which reflect different accretion processes? \citet{Boccardi2021} investigate these aspects using a sample of 27 sources and defining a LERG and a HERG based on the ratio of the x-ray to Eddington luminosity, $L_{x-ray}/L_{Edd}$. Those with a ratio $>1.1\times10^{-3}$ are considered to be HERGs, while those below as LERGs. Although the sample is small and limited by redshift, the HERGs tend to show a transition above $10^6 R_s$ while the LERGs below this limit (Fig.~\ref{f:Boccardi_heg_leg}). This suggests a relationship between jet collimation and the properties of the accretion disk and black hole. \citet{Boccardi2021} also suggest jets in HERGs to have a more prominent outer sheath, and an outer launch radius
$>100 R_s$. Jets in sources such as in M87 appear anchored in the innermost disk regions. BL Lac objects which are the beamed counterparts of FRI sources appear consistent with LERGs in their jet collimation characteristics.

Disk winds may be responsible for the prominent outer sheath in FRII sources and play a prominent role in the collimation of jets. These winds could be probed for example by x-ray detection of ultrafast outflows (UFOs) in AGN originating in the accretion disk \citep{Tombesi2010,Tombesi2014,Reynolds2015}.  Most of the HERGs in the \citet{Boccardi2021} sample exhibit UFOs suggesting that disk winds can be a viable process for the collimation of jets in these sources. As most of the HERGs belong to the FRII category, it is conceivable that the FRII sources have a prominent sheath which stabilises the inner spine and minimises entrainment from the interstellar medium (\citealt{Perucho2006} and references therein). The FRI jets on the other hand are more prone to entrainment and more dissipative. Besides enlarging the sample, it would also be important to study the collimation of jets in FRI HERGs and the FRII LERGs (cf. \citealt{Mingo2022}).

\begin{figure*}[ht!]
    \centering
    \includegraphics[scale=0.125]{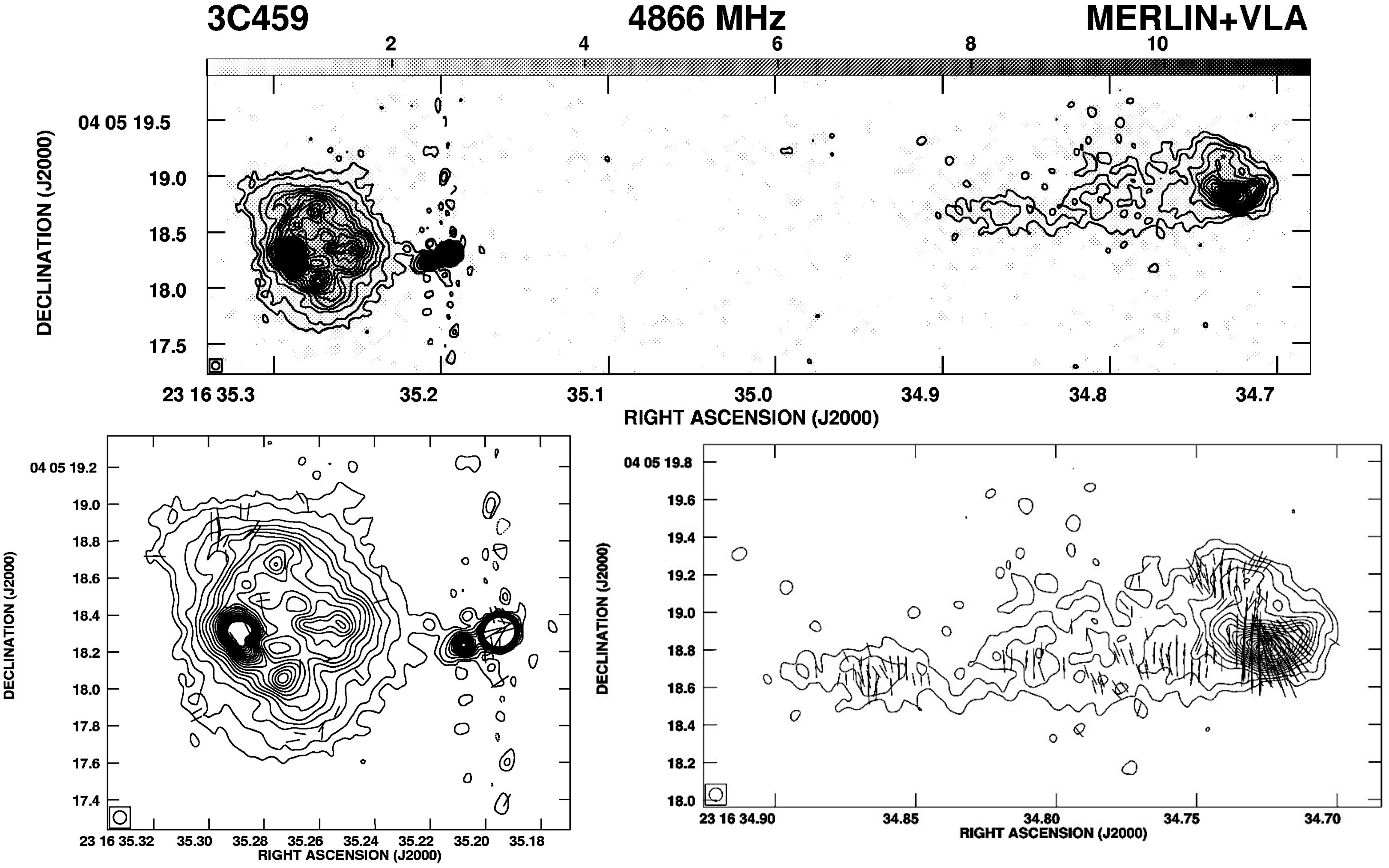}
    \caption{ MERLIN images of 3C459 with an angular resolution of 70 mas. The upper panel shows the total-intensity image, while the
lower panel shows the components with the polarization E-vectors are superimposed on the total intensity contours. The eastern component in this highly asymmetric source is  strongly depolarized possibly due to interaction with the external medium \citep{Thomasson2003}.}
    \label{f:3C459_Thomasson}
\end{figure*}

\section{Jets in Compact steep-spectrum and peaked-spectrum sources}

Our understanding of compact steep-spectrum (CSS) and peaked-spectrum (PS) sources, defined to be less than 20 kpc in size with the latter exhibiting a peak in the radio spectrum, has been reviewed recently by \citet{ODea2021}. The three main scenarios for the nature of CSS and PS sources are as follows. (i) The PS sources are young sources with those which peak at higher frequencies being smaller and younger. The PS sources evolve to the CSS sources which in turn evolve into the larger sources as the jets propagate outwards through the interstellar medium of the host galaxy and later through the intragroup/intracluster medium and then the intergalactic medium. (ii) The jets in CSS and PS sources may be confined to small dimensions within the confines of their host galaxies due to a dense interstellar medium. The jets may also be disrupted. (iii) Alternatively the jets in these sources may be intermittent. Although each of these scenarios may be applicable to different sets of sources, not all CSS and PS sources are likely to evolve into large radio galaxies and quasars. In this Section we summarise a few salient features relevant for the propagation of jets.

The CSS and PS sources are smaller than the dimensions of the host galaxy. Hence the effects of propagation of the jets through the interstellar medium of the host galaxy  
can be probed via both structural and polarization properties of the lobes. Also feedback processes of the jets which may affect the interstellar medium of the host galaxy as well as star formation can be studied from the properties of the host galaxy and the interstellar medium.

\subsection{Jet propagation in an asymmetric environment}
Most of the CSS and PS sources when observed with high angular resolution exhibit a double-lobed structure, often with a radio core particularly in the case of quasars, although the jets in some appear quite distorted and complex. Examples of the latter include 3C48 and 3C119, suggesting disruption of the jet via interaction with the interstellar medium of the host galaxy. For the ones with a double-lobed structure selected largely from strong-source samples, \citet[and references therein]{Saikia2003b} investigated the symmetry parameters of CSS and PS sources such as the separation ratio, flux density ratio of the oppositely directed lobes and the overall misalignment of the sources compared to the larger radio galaxies and quasars. The CSS and PS sources were found to be more asymmetric and misaligned possibly due to interaction of the jets with an asymmetric external environment. The more luminous lobe with the more prominent hotspot being nearer to the core suggests interaction with an external environment rather than this being due to effects of orientation and relativistic motion. In the latter case, the more prominent hotspot would have been on the approaching side of the jet and farther from the nucleus. Similar results were found for weaker source samples as well (\citealt{KunertBajraszewska2016} and references therein).

The external environment which is a magnetoionic plasma through which the jets are propagating can also be probed via polarization observations. The medium will cause a rotation of the E-vector of the synchrotron radiation, the degree of rotation being given by $\chi(\lambda) = \chi_o + RM\lambda^2$ where the rotation measure $RM = 812 \int n_e B_\parallel dl$ rad m$^{-2}$. Here $\chi(\lambda)$ is the position angle (PA) of the E vector at a wavelength $\lambda$,
$\chi_o$ is the PA at zero wavelength, $n_e$ is the electron density in cm$^{-3}$, the parallel component of the magnetic field $B_\parallel$ is in units of mG and $l$ in parsec.
Depolarization of the radio emission can occur due to thermal plasma mixed within the synchrotron emitting region, with the emission from different depths being rotated by different amounts, as well as by unresolved structures in an external medium or screen which may have very different rotation measures.

We need observations of high angular resolution to adequately resolve the structure to probe the environment through which the jets are propagating. \citet{Fanti2001} observed the B3-VLA sample of sources and reported significant asymmetries in polarization of the oppositely directed lobes, again suggesting asymmetries in the external environment. \citet{Saikia2003} studied a stronger source sample of 3CR and 4C sources and also found a significantly higher degree of polarization asymmetry in the lobes, compared with a control sample of larger sources. They argued that this is unlikely to be due to orientation as seen in the Laing-Garrington effect, but reflects an asymmetry in the environment through which the jets are propagating. This is sometimes also seen in radio galaxies and quasars with sizes larger than that of the CSS sources. Two striking examples are two of the most asymmetric sources 3C254 associated with a quasar and 3C459 associated with a galaxy (Fig.~\ref{f:3C459_Thomasson}) although with sizes $>$20 kpc \citep{Thomasson2003,Thomasson2006}. In both cases the lobe located far closer to the nucleus is strongly depolarized relative to the one on the opposite side, suggesting interaction of the jet with a dense magnetoionic cloud.

\begin{figure*}
    \centering
    \includegraphics[width=15cm]{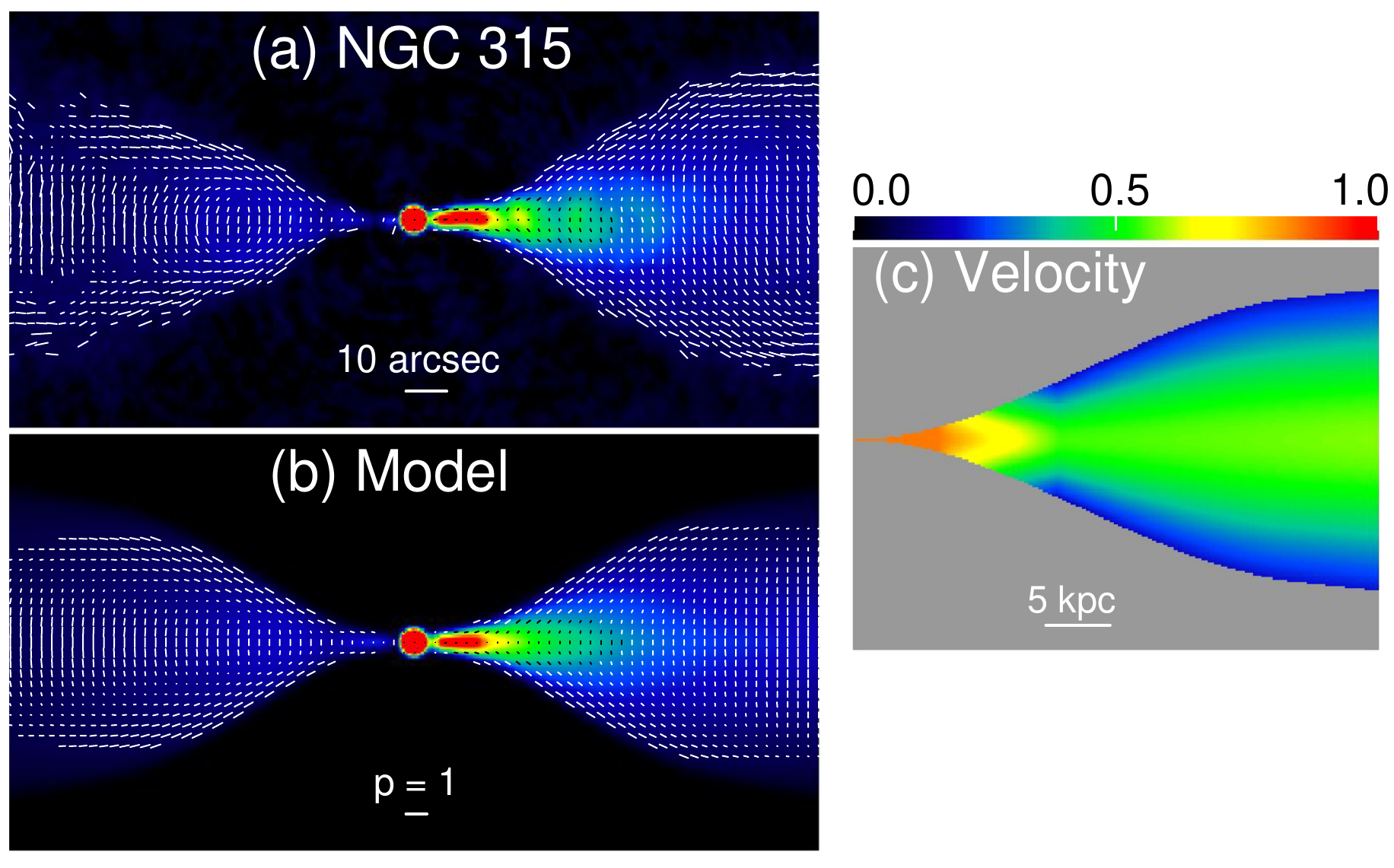}
    \caption{Observations and models of the nearby radio galaxy NGC 315 \citep{Laing2014,Laing2015}. (a) False colour radio image of the galaxy with the vectors
    denoting the degree of polarization, p, and direction of the apparent magnetic
field. (b) Their model fit to the observations shown in (a). (c) The velocity field derived from their model in units of c. Figure is from \citet{Laing2015}. }
    \label{f:NGC315_Laing}
\end{figure*}

The asymmetries in the environment can also be probed via estimates of the rotation measure or $RM$ of the lobes.
Although many CSS sources are known to have high values of $RM$ extending to thousands of rad m$^{-2}$, there are also sources with low values (cf. \citealt{ODea2021} and references therein). Detailed imaging of these sources show that evidence of interaction of jets is quite common. In the CSS quasar 3C147, \citet{Junor1999} reported a huge differential $RM$ with the southern one facing the jet and much closer to the nucleus having a value of $\sim -$3140 rad m$^{-2}$ compared with +630 rad m$^{-2}$ for the more distant one.
VLBI-scale polarimetric observations of the parsec-scale jet suggest $RM$ values ranging from $-1200$ - $-2400$ rad m$^{-2}$ \citep{Zhang2004,Rossetti2009}. \citet{Junor1999} suggested that the jet is interacting with a dense cloud of gas embedded in the magnetoionic medium of the host galaxy, which is hindering its advance. Other examples of enhanced $RM$ near where the jet bends due to interaction with the external environment include B0548+165, B1524$-$136 and 3C119 \citep{Mantovani2002,Mantovani2010}. Sources with a complex morphology possibly due to disruption of the jet may also have a high $RM$ as in 
3C119, 3C318 and 3C343 \citep{Mantovani2005,Mantovani2010}. 
However relationship between the total-intensity structure and $RM$ variations are rich and diverse. For example \citet{Cotton2003} find a high $RM$ and a brightening of the jet in 3C43 where the jet bends, while 3C454 appears to have a high $RM$ across the jet with no significant increase in either brightness or $RM$ where the jet bends. In this case the bend in the direction of the jet may not be due to collision with a cloud of gas.

The possibility of jets propagating through a dense asymmetric medium on opposite sides has also been inferred from measurements of velocities of hotspots from their proper motion as in J0111+3906 and J1944+5448 \citep{Polatidis2002,Rastello2016}. The closer hotspots which are also brighter are moving with smaller velocities. Inferring velocities from estimates of radiative ages of very asymmetric sources also suggest that the nearer components are moving with slower velocities \citep{Orienti2007a}. H{\sc{i}} in absorption has been detected towards the closer and brighter hotspots in the CSS sources 3C49 and 3C268.3, which are also depolarised and associated with optical emission-line gas \citep{Labiano2006}.  

The radio structures of CSS and PS sources are often asymmetric suggesting that the jets are propagating through an asymmetric external environment on these scales with greater dissipation of energy on the jet side \citep{Bicknell2003,Jeyakumar2005}. However in reasonably symmetric sources the hotspots may be traversing outwards with similar velocities as for example in  4C31.04 \citep{Giroletti2003} and J1511+0518 \citep{An2012}.

\section{Large-scale jets}
The large-scale kpc-scale jets in the FRI and FRII sources which can extend up to hundreds of kpc appear structurally different. The FRII jets appear well collimated all along leading to the formation of hot-spots at the lobes, often at the outer edges. The jets in FRI sources tend to `flare' exhibiting an increase in the width of the jet with distance from the AGN. Both FRI and FRII jets appear asymmetric close to the AGN, but the FRI jets are more symmetric on larger scales. The broad picture is that the jets in both FRI and FRII sources are initially relativistic, but the jets in FRI sources decelerate while those in FRII sources remain relativistic till it reaches the outer lobes. The FRI jets are more prone to mass loading or entrainment as it traverses outwards. Mass loading could be due to either stellar winds from within the volume traversed by the jet \citep[e.g.][]{Komissarov1994,Bowman1996} or entrainment of material from the interstellar medium of the host galaxy (e.g. \citealt{Bicknell1986,Rosen2000}, and references therein). 

A number of key and outstanding questions related to our understanding of jets on kpc scales have been highlighted by \citet{Laing2015}. These include the velocity fields of the jets especially the differences in the FRI and FRII sources; jet composition and effects of entrainment; magnetic field structure; confinement of jets and effects of the external environment; generation of relativistic particles and the effects of feedback on the interstellar medium on small scales and on intracluster or intergalactic medium on larger scales. Detailed studies of jets at radio wavelengths have been possible for the FRI sources which can be well resolved transverse to the jet axis, while the jets in FRII sources are narrower. 

\subsection{Jets in FRI sources}
One of the more detailed empirical models to understand jets in FRI radio galaxies was developed by \citet{Laing2002a,Laing2002b} initially applying it to the radio galaxy 3C31. They assume that the two oppositely-directed jets are axisymmetric, intrinsically symmetrical and stationary. The jets are shown to be relativistic so that the apparent asymmetries due to relativistic aberration are much larger than intrinsic asymmetries. They model the jet geometry, three-dimensional distributions of velocity, emissivity and magnetic field structure. These are optimised by comparison with high-resolution images. 

They suggest that the jets can be divided into three parts where the inner region is well-collimated. This is followed by a region of rapid expansion, referred to as the flaring region after which the jets recollimate, and then there is a conical outer region. The magnetic field structure is primarily longitudinal and toroidal. The on-axis velocity drops at the end of the flaring region from $\sim$0.8c to $\sim$0.55c, decreasing further outwards. The velocity at the edges of the jet are significantly lower, with the deceleration of the jet being possibly due to entrainment from the external medium. \citet{Laing2002b} suggest that entrainment from the galactic atmosphere is the dominant process at large distances, while stellar mass loss could make a significant contribution near the flaring point.

This has been extended to a larger sample of FRI sources which have been observed with high sensitivity with the VLA by \citet{Laing2013}, who provide a summary of the results. The observations and model for one of their galaxies, NGC315, are shown in Fig.~\ref{f:NGC315_Laing}.  In the regions where the jets are resolved in the transverse direction, the jets appear to flare increasing in opening angle before recollimating and then having a conical outflow at a distance $r_o$ from the AGN. The velocity is $\sim$0.8c at $\sim 0.1r_o$ where the jet brightens rapidly. The high emissivity   
continues till $\sim 0.3r_o$ with rapid deceleration starting $\sim 0.2r_o$ and continuing till $\sim 0.6r_o$, followed by a constant flow speed. The outflow speed at the jet edges are slower than in the spine of the jet.  The magnetic field is predominantly longitudinal close to the AGN but predominantly toroidal after recollimation. The flaring region would require reacceleration of the ultrarelativistic particles to compensate for the adiabatic losses. Also x-ray synchrotron emission is observed from this region. The evolution and observed characteristics of the jets are best understood in terms of interaction with the external environment, with most entrainment occurring before recollimation \citep{Laing2014,Laing2015}.

High-quality images of eleven FRI jets showed that the spectral index between 1.4 and 8.5 GHz decreases with distance from the nucleus
in all the sources \citep{Laing2013}, similar to what was seen earlier \citep{Laing2008}. The mean spectral index when the jets first brighten abruptly is 0.66$\pm$0.01 and after the jets recollimate the mean spectral index flattens to 0.59$\pm$0.01. The mean change in spectral index which is more robustly measured is $-0.067\pm0.006$ \citep{Laing2013}. Their jet model associate this with a decrease in the jet velocity from $\sim$0.8c to less than $\sim$0.5c, reflecting the particle acceleration processes at play. The possibility of first-order Fermi acceleration would require shocks all along the volume of the jet. The remarkable similarity of the spectral index evolution along the FRI jets studied by \citet{Laing2013}, especially when normalized to the same recollimation distance is striking. This is in contrast to the FRII jets which exhibit greater dispersion in their spectral indices and often have steeper spectra. Although we need to extend such studies to FRII jets and also for a larger sample of FRI jets, the difference suggests that the dominant particle acceleration processes may be different for the FRI and FRII jets.

A detailed study of the total intensity and linear polarization asymmetries of the jets in two FRI galaxies B2~0206+35 (UGC1651) and B2~0755+37 (NGC2484)
has been made by \citet{Laing2012}. They have shown that the asymmetries can
be understood if the jets are intrinsically symmetrical with decelerating relativistic outflows but are also surrounded by mildly relativistic backflows. The backflow velocities are in the range of $0.05<\beta<0.35$ and could be traced to distances from the AGN of at least $\sim$15 kpc and 50 kpc for B2~0206+35 and B2~0755+37 respectively. Backflows are normally associated with FRII sources, and it is interesting to find examples among FRI radio sources. There are a number of open questions listed by \citet{Laing2012}, including where does the backflow start and where does it end and how ubiquitous is it among FRI sources, which the new generation of radio telescopes will help address.

\subsection{Jets in FRII sources}
Detailed modelling as has been done for the extended jets in FRI sources has not been possible for the ones in FRII sources due to inadequate resolution. Also jets in FRII radio galaxies are often quite weak making it difficult for detailed work with the current sensitivity limits. 

Collimation of jets has been probed by examining the structure of jets, especially in quasars and possible correlation of the size of hotspots with projected linear size. In their detailed study of 12 3CR quasars, \citet{Bridle1994} examined the variation of width transverse to the jet axis with distance from the core, and found that after an initial rapid expansion, the expansion slows down and there is evidence of recollimation. This is consistent with studies of other jets as well. However they note that while the spreading rate, defined as the ratio of knot width to distance from core, is often $>$0.1 for jets in low-luminosity sources as seen for the FRI sources, but rarely $>$0.1 for high-luminosity ones. Collimation of jets in FRII sources may also be examined by studying the variation of hotspot size with projected linear size.  For a sample of FRII radio galaxies larger than about 70 kpc, \citet{Hardcastle1998c} found the hotspot size to be correlated with the projected size with a slope of about unity; consistent with the trend noted by \citet{Bridle1994} that the hotspot size scales with linear size. \citet{Jeyakumar2000}
extended this to compact steep-spectrum and peaked spectrum radio sources, defined to be less than about 20kpc, and found that the hotspot size for CSS and PS sources increases with linear size, with some evidence of flattening beyond this scale. The jets in quasars exhibit a significant trend to point towards the more prominent hotspot, while this trend is weaker in the case of radio galaxies (cf. \citealt{Hardcastle1998c}). This must be due to mild relativistic beaming of the hotspots and would be consistent with the unified scheme for radio galaxies and quasars. The high detection rate of jets in quasars compared with a much lower fraction for radio galaxies \citep[e.g.][]{Fernini1993,Hardcastle1998c} is also consistent with the unified scheme.

With what velocities are the jets traversing outwards in the FRII sources? The correlation of jet-sidedness with lobe depolarization, the Laing-Garrington effect \citep{Laing1988,Garrington1988}, demonstrates that the prominent jets are on the approaching side. This is consistent with relativistic beaming being a viable explanation of jet asymmetry. Superluminal motion of nuclear jets show that the nuclear jets are travelling close to the velocity of light. Assuming that the nuclear jets have a typical Lorentz factor $\gamma \sim$5, \citet{Bridle1994} estimate the Lorentz factor of the extended quasar jets in their sample $\gamma_j$ to be $1.6\pm0.2$. This suggests that although extended radio jets may start off with highly relativistic velocities, the velocities on average slow down with increasing distance from the nucleus.
If the jets and their environments were intrinsically symmetric, it would have been in principle possible to estimate jet velocities from the observed asymmetry of the jet to counter-jet brightness or flux density ratio. However, although relativistic beaming appears to play a dominant role, there has also been evidence of intrinsic asymmetries playing a role. 

Examples of this are seen in the form of increased brightness in jets in regions where jets appear to bend. Also, in a small number of cases jets contribute over about 30 per cent of the total flux density of the source in sources with relatively weak cores, and hence likely to be inclined at large angles to the line of sight. Examples include the quasars 3C9 \citep{Bridle1994}, 3C280.1 \citep{Swarup1982} and B1857+566 \cite{Saikia1983}. These jets point in the direction of a weak hotspot, suggesting that the observed jet asymmetry could also be contributed significantly by dissipation of energy in the jet. Examples of weak-cored, one-sided radio sources also suggest intrinsic jet asymmetries in these sources \cite{Saikia1989}, although deeper observations are required to clarify the situation.

\subsection{Transition cases}
In addition to the jets in FRI and FRII sources, also referred to as weak and strong-flavour jets, there are
a number of transition cases which often occur in sources classified as FRI-II. These jets may flare, but do not appear to decelerate significantly and have detected counter-jets \citep{Laing2015}. One of the well-studied examples in this category is NGC6251 which is a giant radio galaxy with an overall size of $\sim$1.56 Mpc, and with a large side-to-side ratio for the oppositely-directed jets (\citep{Perley1984,Laing2015}.  

\subsection{Giant radio sources}
Although giant radio sources (GRSs) have traditionally been defined to be $>$1 Mpc  \citep[e.g.][]{Schoenmakers2001}, a limit of 0.7 Mpc has been widely used recently with the current cosmological parameters (e.g. \citealt{Kuzmicz2018,Dabhade2020a,Dabhade2020b}). Among the giant radio sources, most of them belong to the FRII class, with some in the intermediate FRI/II category and only a small fraction in the FRI class. The FRIs include tailed radio sources in clusters of galaxies. such as 3CR129 \citep[e.g.][]{Lane2002,Lal2004} and 3CR130 \citep[e.g.][]{Hardcastle1998b}. In the early compilation by \citet{Ishwara-Chandra1999}, of the 53 GRSs, only 4 were classified as FRIs. The percentage of FRIs in the LoTSS \citep{Dabhade2020a} and SAGAN samples \citep{Dabhade2020b} are similar. In the compilation by \citet{Kuzmicz2018}, of the 349 GRSs only 20 are FRIs, again a similar percentage to that of \citet{Ishwara-Chandra1999}. In the \citet{Kuzmicz2018} sample, the median projected size of the FRI GRSs was found to be lower than those of FRIIs. The FRIs also appear confined to the nearby Universe with a maximum redshift of $\sim$0.24, while the median redshift of the entire sample is 0.24 with the highest value being 3.22 and 28 objects having a redshift $>$1.
As the jets in FRI sources are highly dissipative as they traverse outwards, it is possible that a significantly smaller fraction may be able to reach sizes $>$0.7 Mpc compared with FRIIs. However, the small fraction may also be partly due to difficulty in detecting weak diffuse emission, especially at high-redshifts where inverse-Compton losses with the cosmic microwave background may dominate over synchrotron losses. Deep radio observations sensitive to diffuse large-scale structure with the required resolution at low radio frequencies should help clarify whether FRI GRSs may be more common than presently observed. In the nearby FRI radio galaxy 3CR31, deep low-frequency observations have revealed that the plumes of radio emission extend much farther than earlier seen making the projected linear size $\sim$1.1 Mpc \citep{Heesen2018}. Modelling the jets in FRIs from high-resolution observations has been discussed in Section 7.1, where modelling of the  jets in the giant radio galaxy NGC315 \citep{Laing2014,Laing2015} has also been discussed. The jets in giant radio galaxy 3CR31 have been modelled with an inclination angle of $\sim$52$^\circ$ to the line of
sight, an on-axis jet speed $\beta\sim$0.9 at 1 kpc from the nucleus, decelerating to $\sim$0.22 at 12 kpc, with slower speeds at the edges \citep{Laing2002a,Laing2002b}. Deriving an external pressure profile from x-ray observations \citet{Croston2014} extend the modelling of entrainment to $\sim$120 kpc. \citet{Heesen2018} have extended the analysis to larger distances.  

The fraction of sources with well-defined jets in FRII GRSs is small \citep{Dabhade2020b,Kuzmicz2021}. In a sample of 174 GRQs, less than $\sim$3 per cent have been found to exhibit radio jets \citet{Kuzmicz2021}. As GRSs may on average be expected to be inclined at larger angles to the line of sight than smaller sources associated with similar hosts, jets may appear weaker due to Doppler effects. However, the quasars are expected to be inclined within $\sim$45$^\circ$ to the line-of-sight in the unified scheme for FRII radio galaxies and quasars \citep[e.g.][]{Barthel1989}, and the relative core strengths of giant radio quasars (GRQs) were found to be broadly consistent with the unified scheme (cf. \citealt{Ishwara-Chandra1999}). Therefore the extremely low fraction of quasars with radio jets is somewhat surprising, and deeper observations with adequate resolution should help clarify this aspect. Our current understanding of GRSs and possibility of future studies with SKA are being discussed by \citet{Dabhade2022}.

\subsection{Jets in tailed radio sources}
Extragalactic radio sources with a head-tail shape where the parent optical galaxy is at the head of the radio source were first identified by \citet{Ryle1968}. More examples of such sources led \citet{Miley1972} to suggest that the structure of these sources is due to bending of the jets by ram pressure of the intracluster medium. Further observations of radio sources in clusters of galaxies showed that the opening angle of the tails had a large range, those with small opening angles were termed narrow-angle tailed (NAT) sources, while those with larger opening angles as wide-angle tailed (WAT) sources. The WATs appear to be of higher radio luminosity and tend to be associated with the dominant galaxies in clusters, although not exclusively \citep[e.g.][]{Owen1976,GoldenMarx2019}. 

Clusters of galaxies are dynamical systems where there could be infall of individual galaxies or small groups, and mergers of clusters of similar mass or a smaller cluster merging with a bigger one. These interactions lead to the development of turbulence, shocks and sloshing motions in the intracluster medium (ICM). In addition there is feedback from AGN in clusters of galaxies, with the jets often showing evidence of recurrent activity (Section 9). Extended radio sources in clusters are evolving in such a turbulent medium. While the radio jets in NATs may be bent by ram pressure due to motion of the parent galaxy through the ICM, such an explanation is unlikely for WATs as the host galaxies are often the dominant galaxies in clusters and not expected to have high velocities relative to the ICM. The wide range of shapes of the jets and tails is likely due a combination of motion of the galaxy as well as dynamics of the ICM. Their appearance will also be strongly influenced by projection effects. It is also interesting to note that recent observations of tailed radio sources in clusters have revealed new features which pose interesting challenges in understanding these features (e.g. \citealt{Gendron2021} and references therein).

In this article we highlight a few aspects of jets in tailed radio sources in clusters. The jets in NATs are similar to those of FRI sources, except for being bent into a C- or V-shape. Among the archetypal NATs is 3CR83.1B (NGC1265) which was highlighted by \citet{Ryle1968} and \citet{Miley1972}, and studied in detail by \citet{ODea1986,ODea1987}, and more recently by a number of other authors (e.g. \citealt{Gendron2020,Gendron2021}, and references therein). With the jets being swept backwards, the inner jet knots appear brightest in the leading or 'front' edge with higher fractional polarization, and magnetic field along the jet axis. The field lines have been possibly sheared tangentially. In the latter one-third of the jet the magnetic field lines are more complex with a significant perpendicular component. The jets exhibit regions of faster and slower expansion with distance from the core, and wiggles as it traverses outwards possibly due to the development of Kelvin-Helmholtz instabilities \citep{ODea1986,ODea1987}.  

\begin{figure*}
    \centering
    \includegraphics[width=14.0cm]{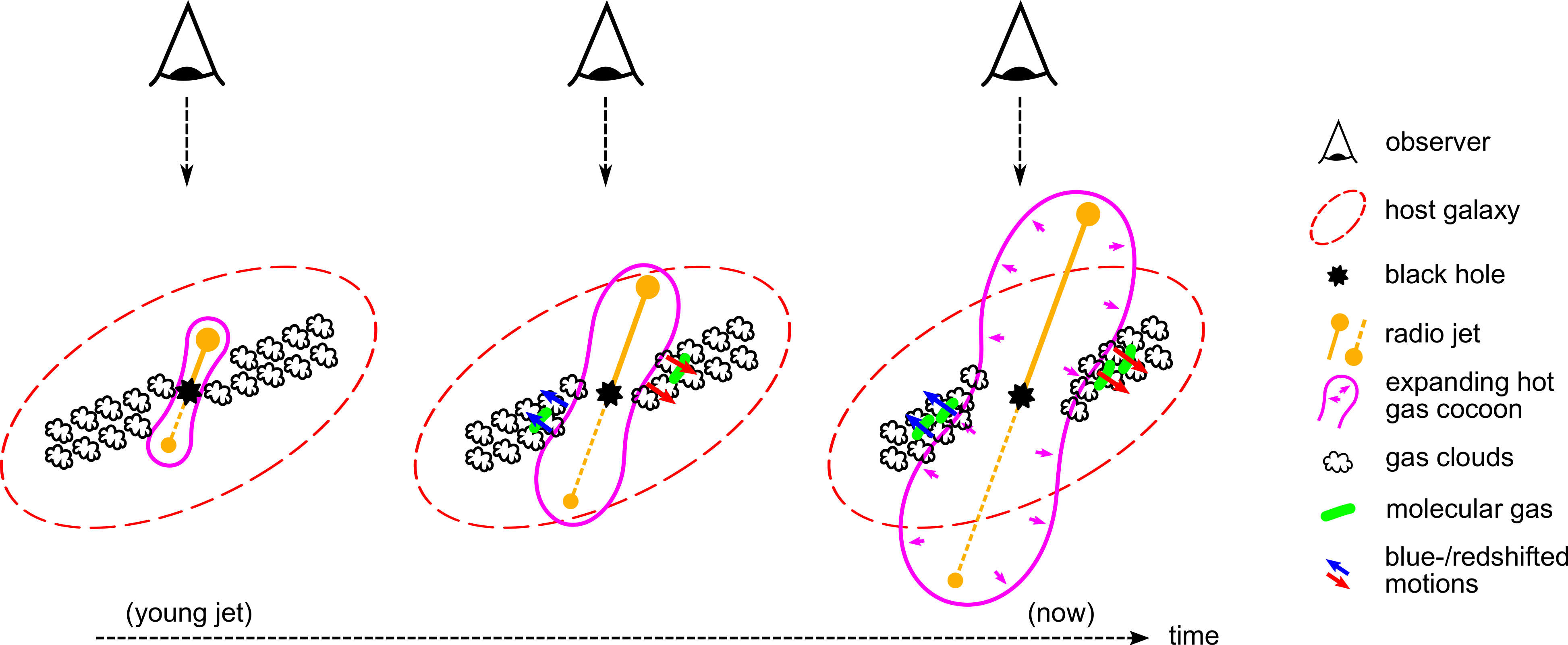}
    \caption{Schematic illustration to explain the gas kinematics in the central region of the quasar 3C273 where a rotating gas disk is affected by the emerging jet with its associated expanding hot gas cocoon \citep{Husemann2019}. \copyright AAS. Reproduced with permission.}
    \label{f:3C273_Husemann}
\end{figure*}

 One may enquire whether jets and tails which appear unresolved and occur on only one side of a galaxy may represent truly one-sided jets. High-resolution observations of a sample of such sources have shown that almost all have two-sided jets when observed with sufficiently high angular resolution. This demonstrates that these are similar to other NATs. The jet to counter-jet brightness ratio suggests that the large-scale jets are at best mildly relativistic with velocities of $\sim$0.2c, similar to those of FRI radio galaxies \citep{TernideGregory2017}. The velocities could be larger for the nuclear jets. IC310, associated with an SO galaxy appear to have a one-sided parsec scale jet in the observations of \citet{TernideGregory2017}, but was later shown to have two-sided jets on a larger scale \citep{Gendron2020}. It has the most prominent core among the sources observed by \citet{TernideGregory2017}, exhibits blazar-like characteristics \citep[cf.][]{Glawion2017}  and it is likely that in such cases relativistic beaming effects are playing a role. 

The WATs on the other hand have luminosities near or above the classical Fanaroff-Riley break, and when observed with high angular resolution exhibit one or two jets which are well-collimated for tens of kpc and appear similar to those of FRII sources. They then broaden and flare dramatically to form extended plumes or lobes of emission \citep[e.g.][]{ODonoghue1993,Hardcastle1998b}. The large-scale jet velocities have again been found to be mildly relativistic with velocities of $\sim$0.2c from jet to counter-jet brightness ratios. The plumes and lobes of emission were generally seen to bend in the same direction forming a C-shaped structure, but deeper observations may reveal more complex structures reflecting the complexity of the cluster and its ICM. 

For example in the tailed galaxy NGC1272, the collimated jets ``initially bend to the west, and
then transition eastward into faint, 60 kpc long extensions with eddy-like structures and filaments'' \citep{Gendron2020} They suggest that gas motion of the ICM, motion of the galaxy in the cluster including through a sloshing cold front all play a role. More sensitive observations especially with SKA are likely to reveal more such structures and provide deeper insights into the ICM and interaction of the jets with it.  

\section{Jet interaction and feedback}
Feedback processes in AGN include the effects of jets, winds, cosmic rays and radiation on the host galaxy, its interstellar medium and the environment. Among these, the effects of radio jets are perhaps better understood, although it is often difficult to disentangle the different contributions. The energy input from radio jets could regulate star formation suppressing star formation in the massive galaxies and determining the high-mass end of the galaxy luminosity function; prevent cooling flows in clusters of galaxies and help understand the balance of heating and cooling processes in the intracluster medium (cf. \citealt{Benson2003,Croton2006,Fabian2012,McNamara2007,McNamara2012}. Feedback has been invoked to understand the strong colour bi-modality of galaxies suppressing star formation as galaxies move to the red sequence (e.g. \citealt{Baldry2004}, and references therein),
galaxy black-hole and bulge mass correlation \citep{Silk1998}, properties of the circumgalactic medium and evolution of gas in dark matter halos. Radio galaxies and feedback from AGN jets have been reviewed extensively relatively recently by \citet{Hardcastle2020}, and effects on the cold components of the interstellar medium, neutral atomic hydrogen and molecular gas have been reviewed by \citet{Morganti2018}, \citet{Morganti2021} and \citet{Veilleux2020}.

\begin{figure*}
    \centering
    \includegraphics[width=16.0cm]{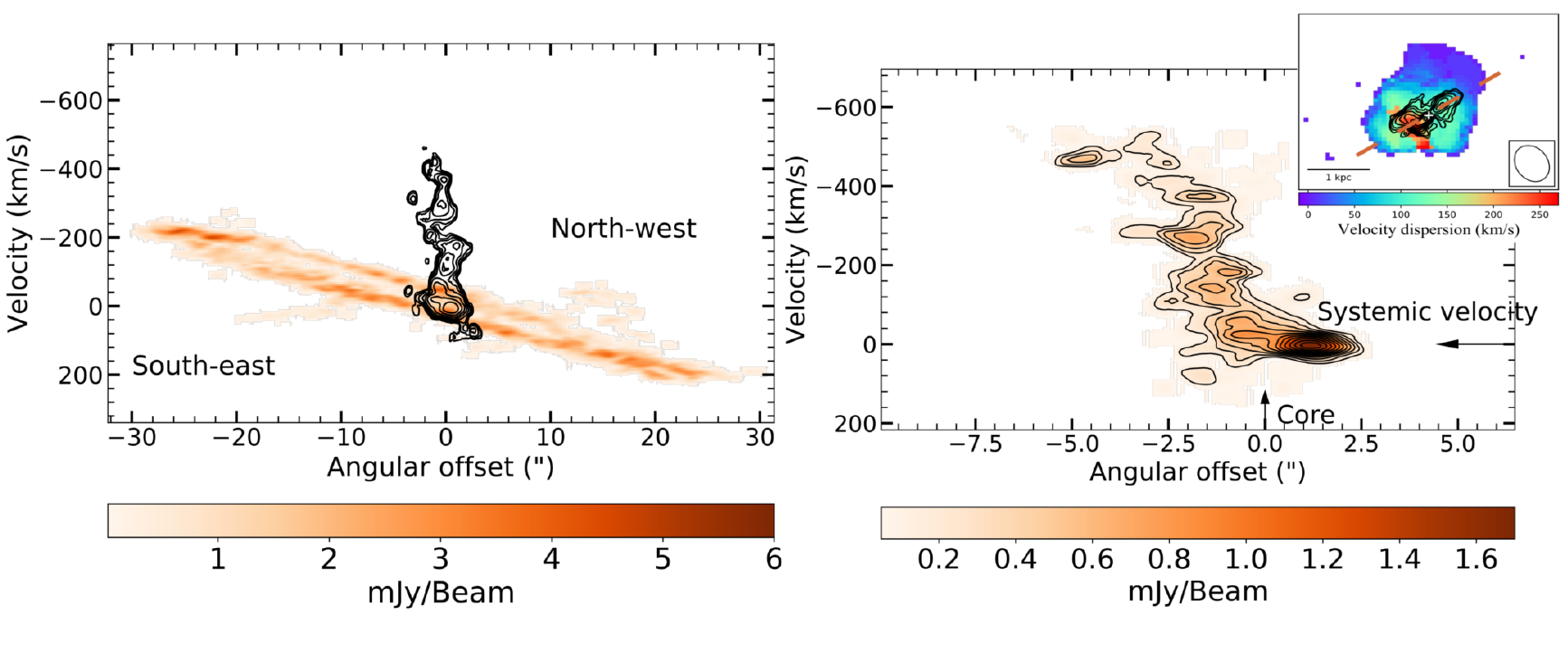}
    \caption{Left panel: The position-velocity (PV) diagram of the large-scale gas disk in colour and the circumnuclear gas in contours in the radio source B2~0258+35 associated with the galaxy NGC~1167. These are from along the major axis of the large-scale gas disk which highlights the different kinematics of the regularly rotating gas and the disturbed gas. Right panel: PV diagram of the circumnuclear gas extracted from along the radio axis \citep{Murthy2022}.}
    \label{f:Murthy_CO_gas}
\end{figure*}

\subsection{The alignment effect and jet-triggered star formation}
One of the early suggestions of jet-cloud interactions triggering star formation was the alignment effect discovered by \citet{McCarthy1987} and \citet{Chambers1987}. Here the optical images of high-redshift radio galaxies appear to align well with the radio axes of the double-lobed radio sources. This was also clearly demonstrated in a sample of 3CR radio galaxies in the  redshift interval $0.6<z<1.3$ observed with the HST, VLA and UKIRT \citep{Best1996}. The elliptical galaxy with its old stellar population was seen in the infrared images, distinctly different from the aligned structures seen at optical wavelengths. The alignment effect evolved with size in this redshift range, being less prominent for larger sources, suggesting that it is a relatively short-lived phenomenon. It was natural to assume that this may be due to formation of young stars triggered by the jet which have evolved on time scales of $\approx 10^7$ yr, similar to those of the evolution of the double-lobed radio galaxies. \citet{Rees1989} developed a model where cold clouds of $\approx$10$^4$K are compressed triggering star formation, while \citet{DeYoung1989} performed numerical simulations to suggest high star formation behind the shock wave as the gas cools. Although attractive, the explanation may be more complex \citep[e.g.][]{Longair1995,Best2000}. Detection of polarized emission suggested that some of it could be scattered light from a hidden nucleus. \citet{Best2000} showed in the young sources the bow shock affects the morphology, kinematics and ionization properties of the emission line gas, while these are more settled for the larger sources. 

Although exploring the alignment effect in GPS sources is a challenge because of their small sizes, it has been done for a couple of GPS sources and a number of CSS objects (see Section 5.5.2 in \citealt{ODea2021} for a summary). The alignment effect in CSS sources is seen at all redshifts although for the larger sources it is confined to $z>0.6$ \citep[e.g.][]{Privon2008}. Recently, \citet{Duggal2021} have
reported extended UV emission co-spatial with the radio source, and have suggested that this may be due to star-formation triggered by the radio jet. Although this remains a possibility which needs further exploration, the alignment effect is perhaps due to a combination of scattered AGN light, nebular continuum emission and star formation. In this context it is also relevant to note that \citet{Collet2015} reported the detection of extended warm ionized gas in two high-redshift galaxies which does not appear to be related to the radio jets, unlike most high-redshift radio galaxies.
They suggest that the extended line emission in these two cases may arise from extended gas disks or filaments in the vicinity of the radio galaxy.

There are a number of well-studied examples of jet-induced star formation in luminous radio galaxies as the jets propagate through the ISM \citep[e.g.][]{Fragile2017}.  These include  Minkowski's Object (\citealt{Zovaro2020}, and references therein), Centaurus A (\citealt{Salome2017}, and references therein), 3C285 
\citep{Salome2015}, 4C 41.17 \citep{Nesvadba2020}, and 3C441 \citep{Lacy1998}. In the `radio-quiet' quasar J1316+1753 the close alignment of the jet and the position angle of the stellar bulge is also suggestive of star formation triggered by the jet which contributes to the stellar bulge (\citealt{Girdhar2022}; see Section 8.2)

\subsection{Suppression of star formation}
Since the early evidence of radio jets affecting the observed properties of the narrow-line regions in samples of Seyfert galaxies \citep[e.g.][]{Whittle1992}, as well as in detailed studies of individual sources such as for example NGC~1068 \citep{Axon1998} and Mkn~3 \citep{Capetti1999}, examples of such interaction have also been found in other low-luminosity AGN \citep[e.g.][]{May2018}. In the nearby radio galaxy Coma A the radio emitting plasma appears closely related to the ionized gas \citep{Tadhunter2000}. The effects of jet-ISM interactions for the luminous compact steep-spectrum and peaked-spectrum radio sources at different wavelengths have been summarized by \citet{ODea2021}. A number of authors \citep[e.g.][]{Morganti2018,Morganti2021,Ruffa2022,Girdhar2022,Murthy2022}, and references therein, have discussed several examples of jet-ISM interaction inferred from from H{\sc i} and CO observations.

In this short review we highlight a few illustrative examples of jet-ISM interactions rather than provide an exhaustive list. Blue-shifted H{\sc i} absorption profiles suggest velocities ranging from a few hundred to $\sim1300$ km s$^{-1}$, mass of a few times 10$^6$
to 10$^7$ M$_\odot$ and outflow rates of about 20 - 50 M$_\odot$ yr$^{-1}$ \citep{Morganti2021}. Noted examples where the absorbing H{\sc i} clouds have been localised include
the restarted radio galaxies 3C293 \citep{Mahony2016} and 3C236 \citep{Schulz2018}, and the CSS source 4C12.50 \citep{Morganti2013}. 
The CSS source 4C31.04 exhibits shocked molecular and ionized gas due to jet-driven feedback 
\citep{Zovaro2019}. They suggest that dense clumps of gas inhibit the advancement of the brightest radio synchrotron emitting plasma, while the less dense material percolate through the porous ISM of the host galaxy. \citet{Nesvadba2010} find most of the molecular gas in the giant radio galaxy 3C326N to be warm, and suggest that a fraction of the mechanical energy of the jet is deposited in the ISM, which provide energy for the outflow besides heating the ISM. Optical observations suggest a mass outflow rate of 30-40 M$_\odot$ yr$^{-1}$
with a terminal velocity of $\sim -1800$ km s$^{-1}$.
Atacama Large Millimeter Array (ALMA) CO(1-0) observations of the giant radio galaxy associated with a spiral host, J2345$-$0449 indicate high gas motions possibly due to the jet kinetic energy \citep{Nesvadba2021}. \citet{Husemann2019} have observed the hyperluminous quasar 3C273 with VLT-MUSE optical 3D spectroscopy and ALMA and find that both the ionized gas in the narrow-line region and the molecular gas are kinematically disturbed. They propose a scenario in which a hot gas cocoon associated with the emerging jet affect the gaseous components in a rotating disk (Fig.~\ref{f:3C273_Husemann}).

\begin{figure*}
    \centering
    \vbox{
       \hbox{
       \includegraphics[width=5.6cm]{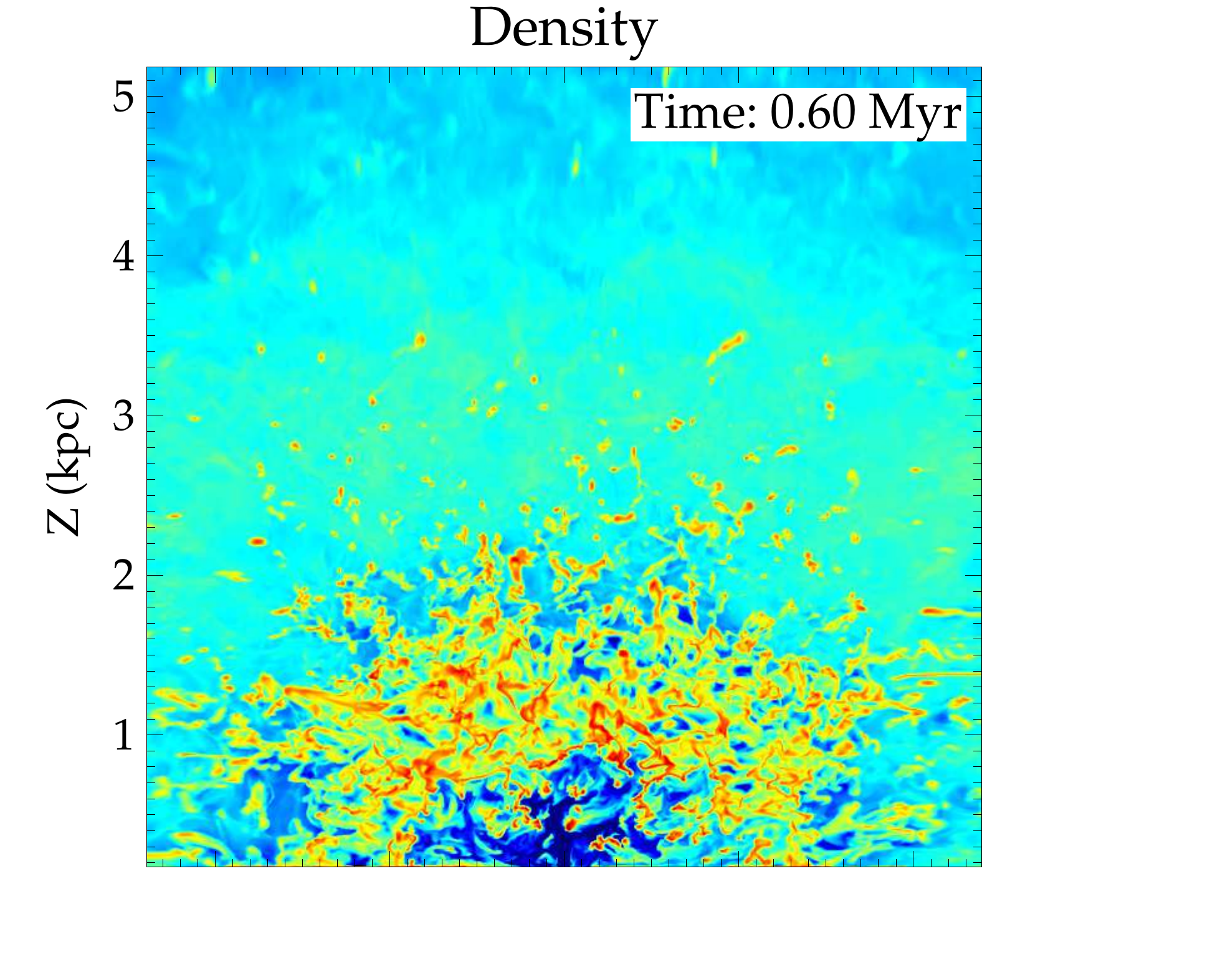}
       \includegraphics[width=5.6cm]{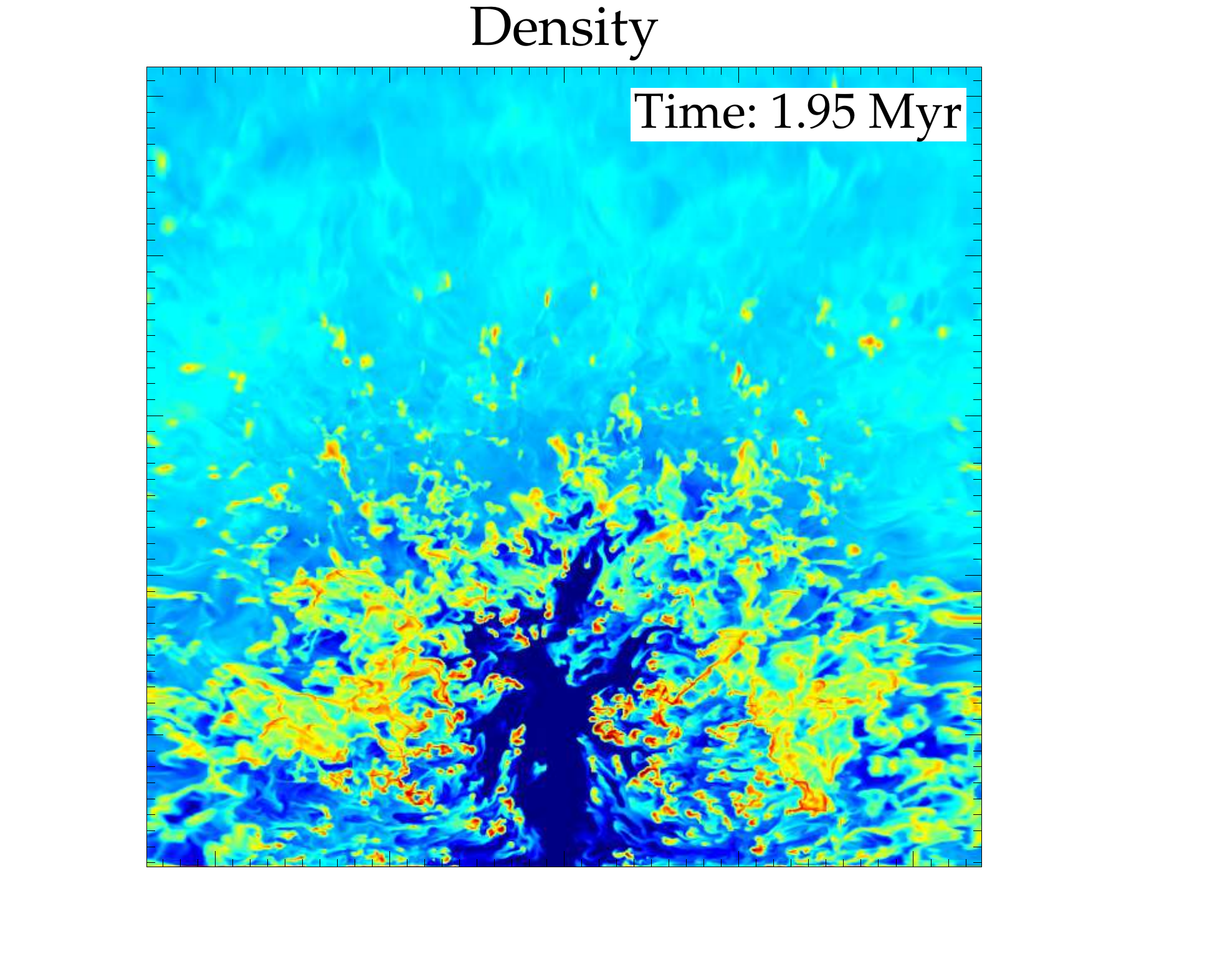}
       \includegraphics[width=5.6cm]{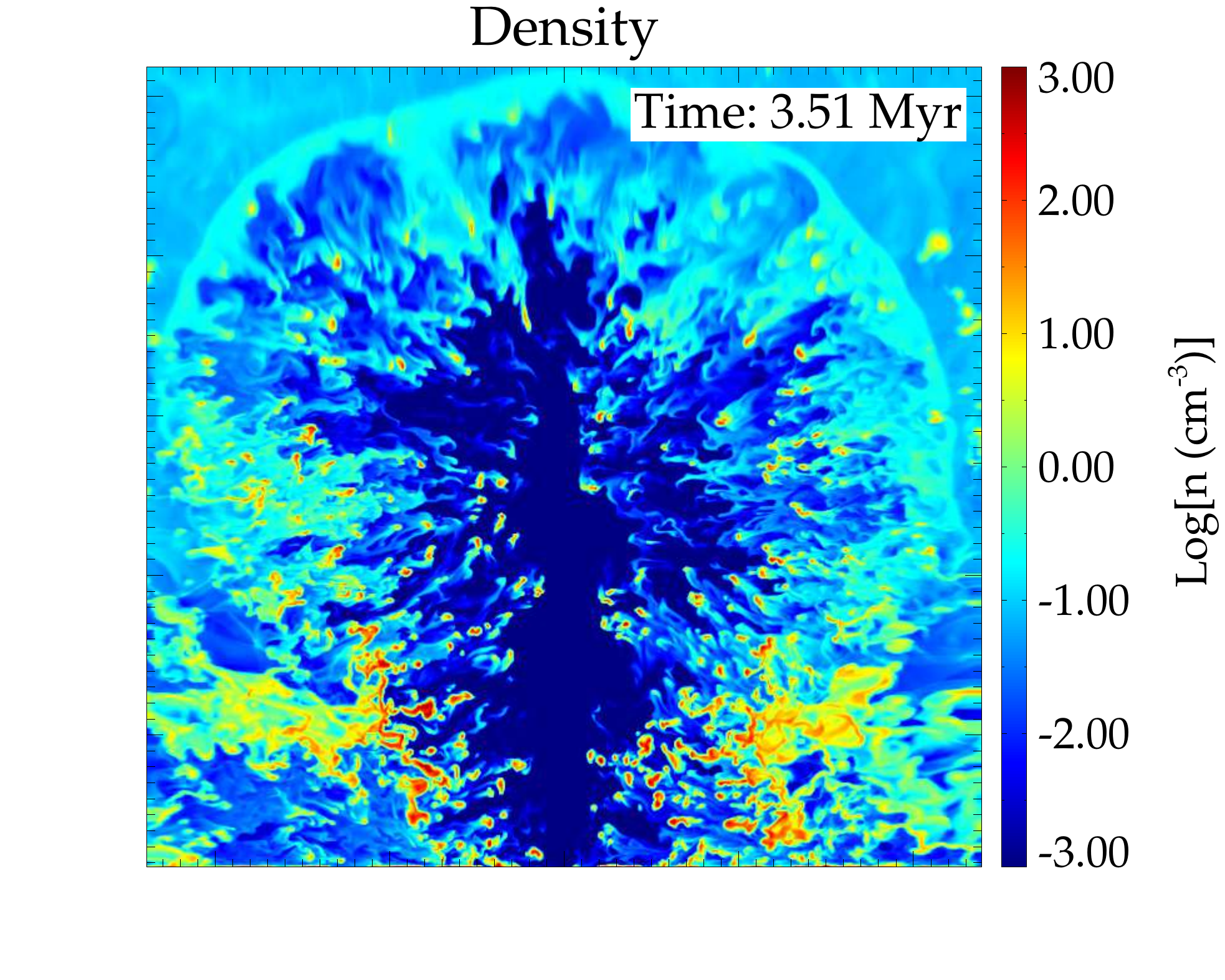}
            }
       \hbox{
       \includegraphics[width=5.6cm]{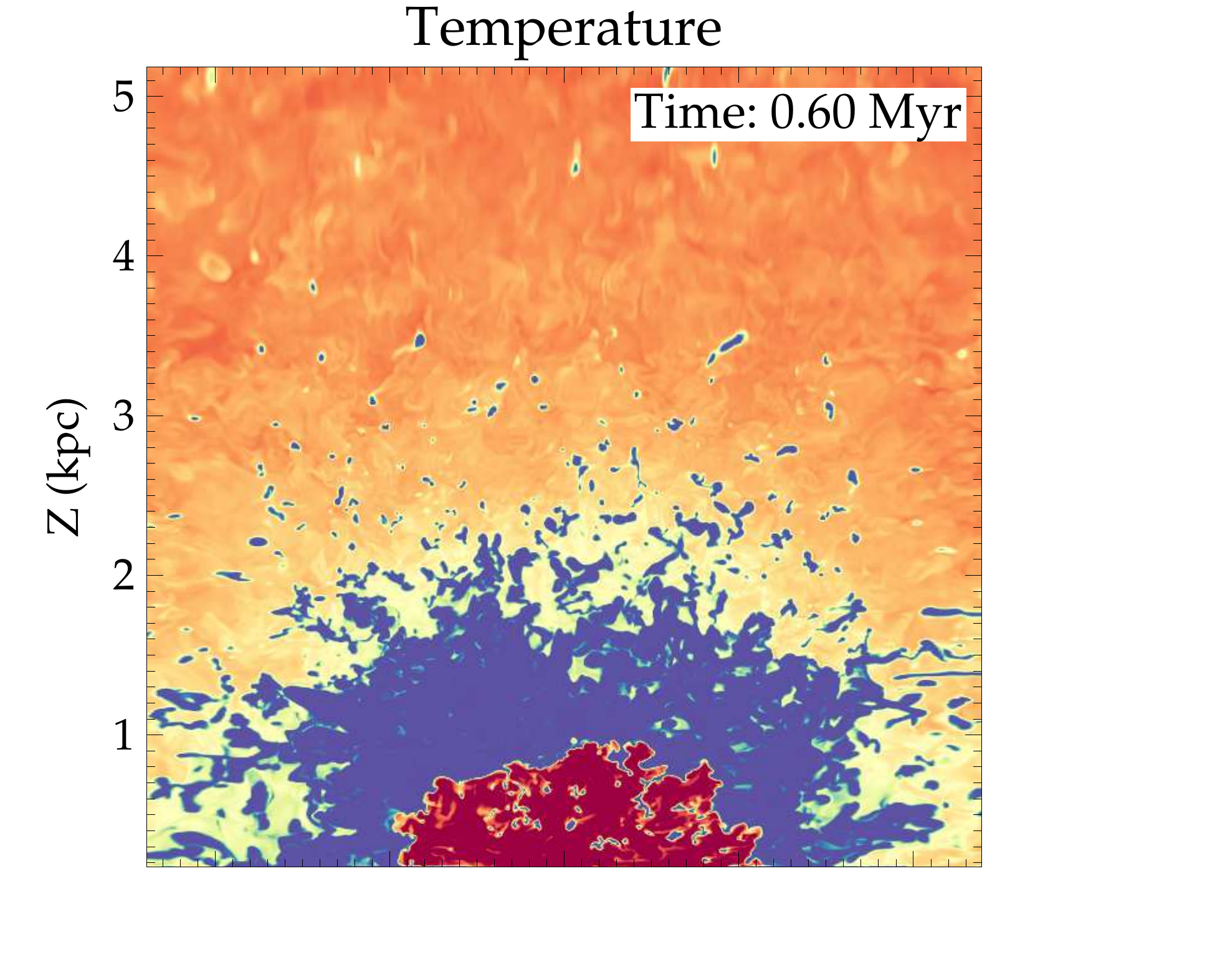}
       \includegraphics[width=5.6cm]{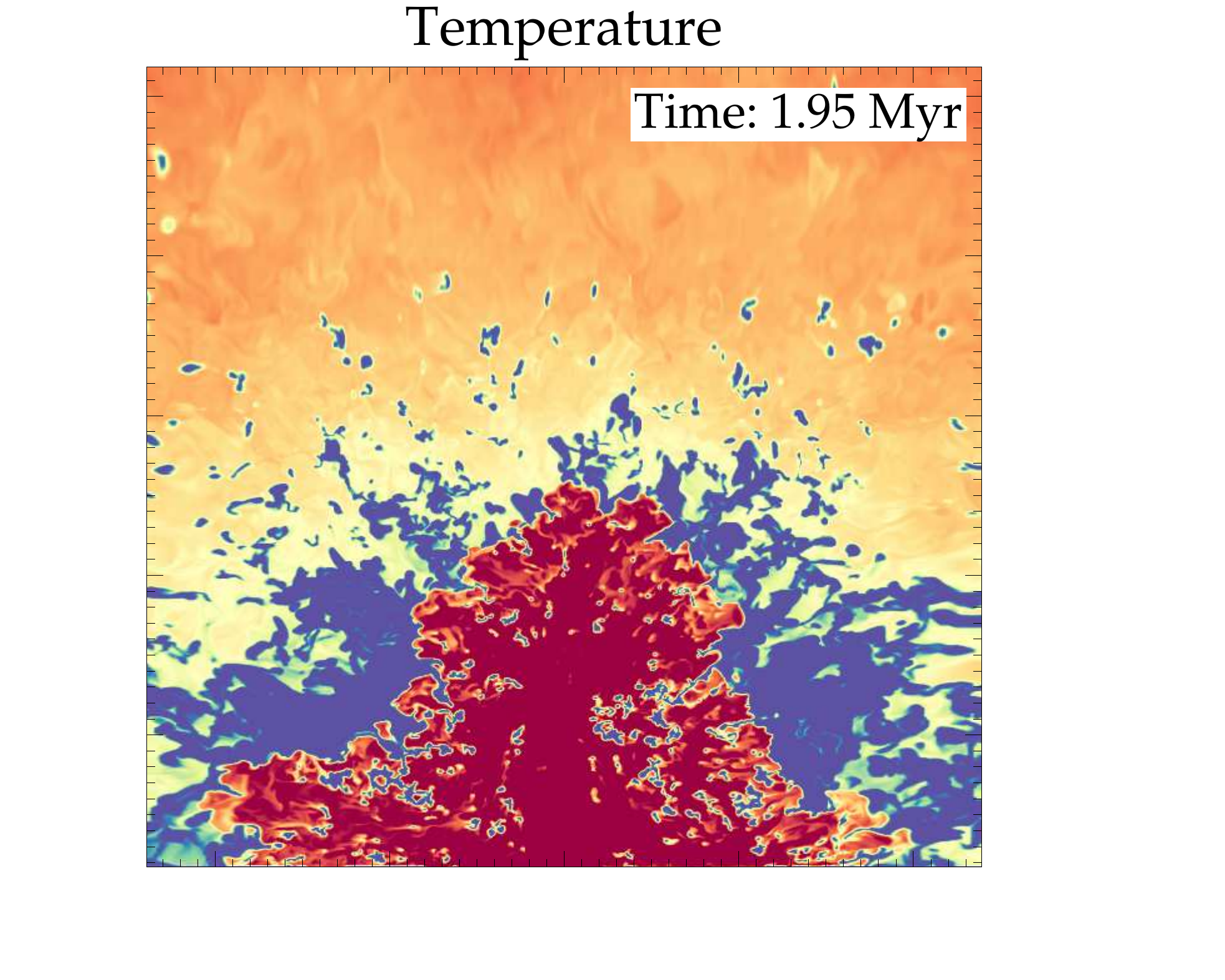}
       \includegraphics[width=5.6cm]{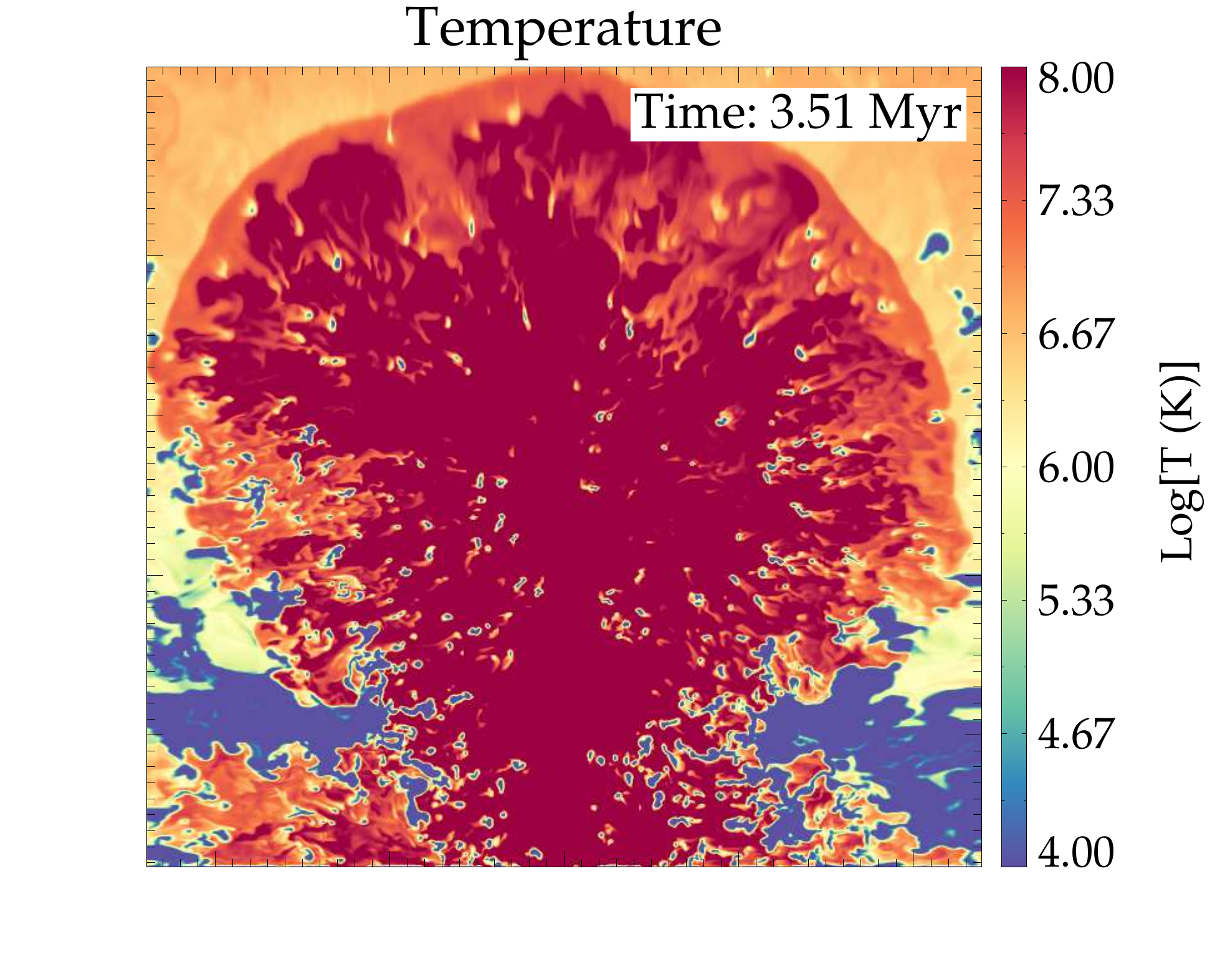}
            }  
        \hbox{
        \includegraphics[width=5.6cm]{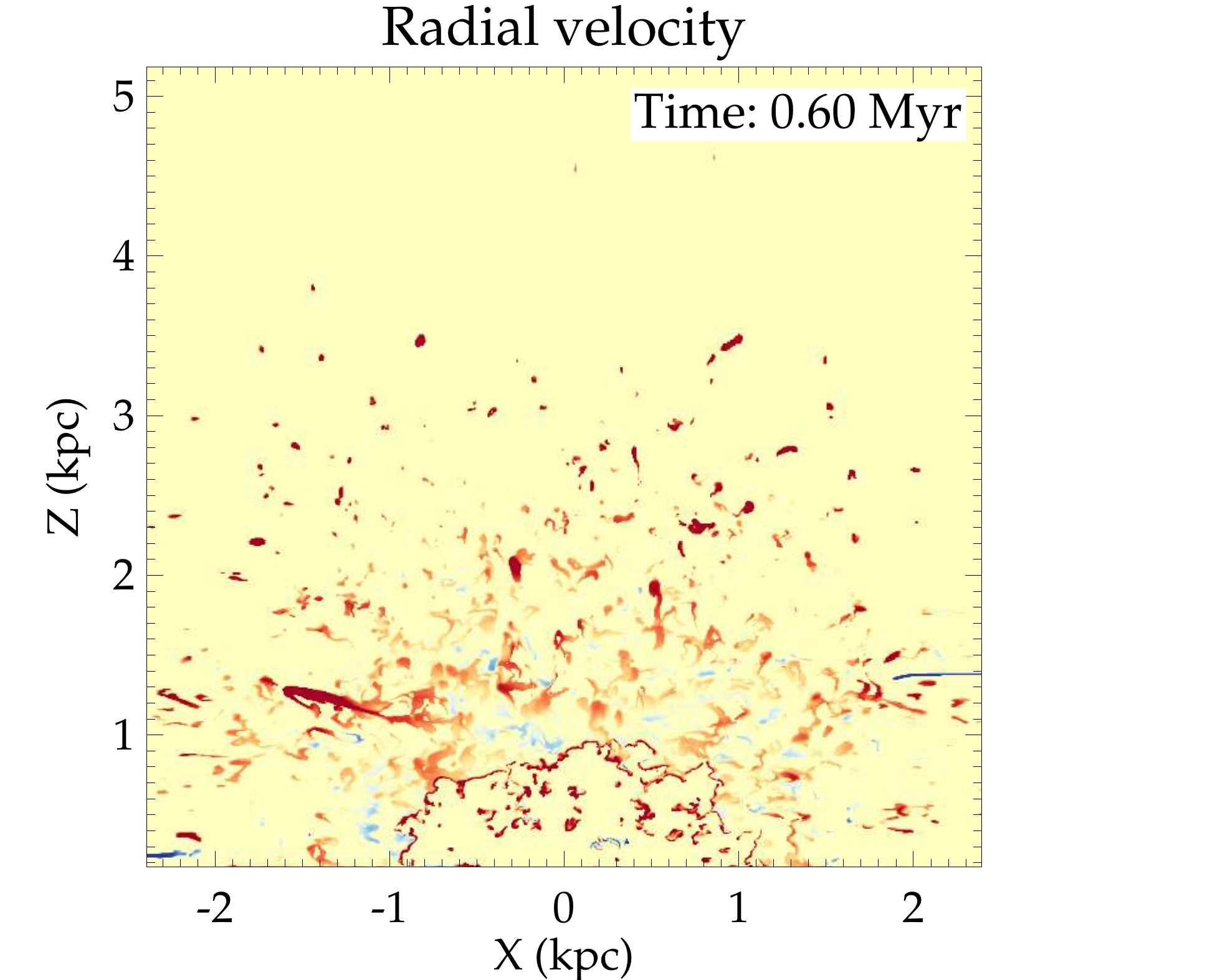}
        \includegraphics[width=5.6cm]{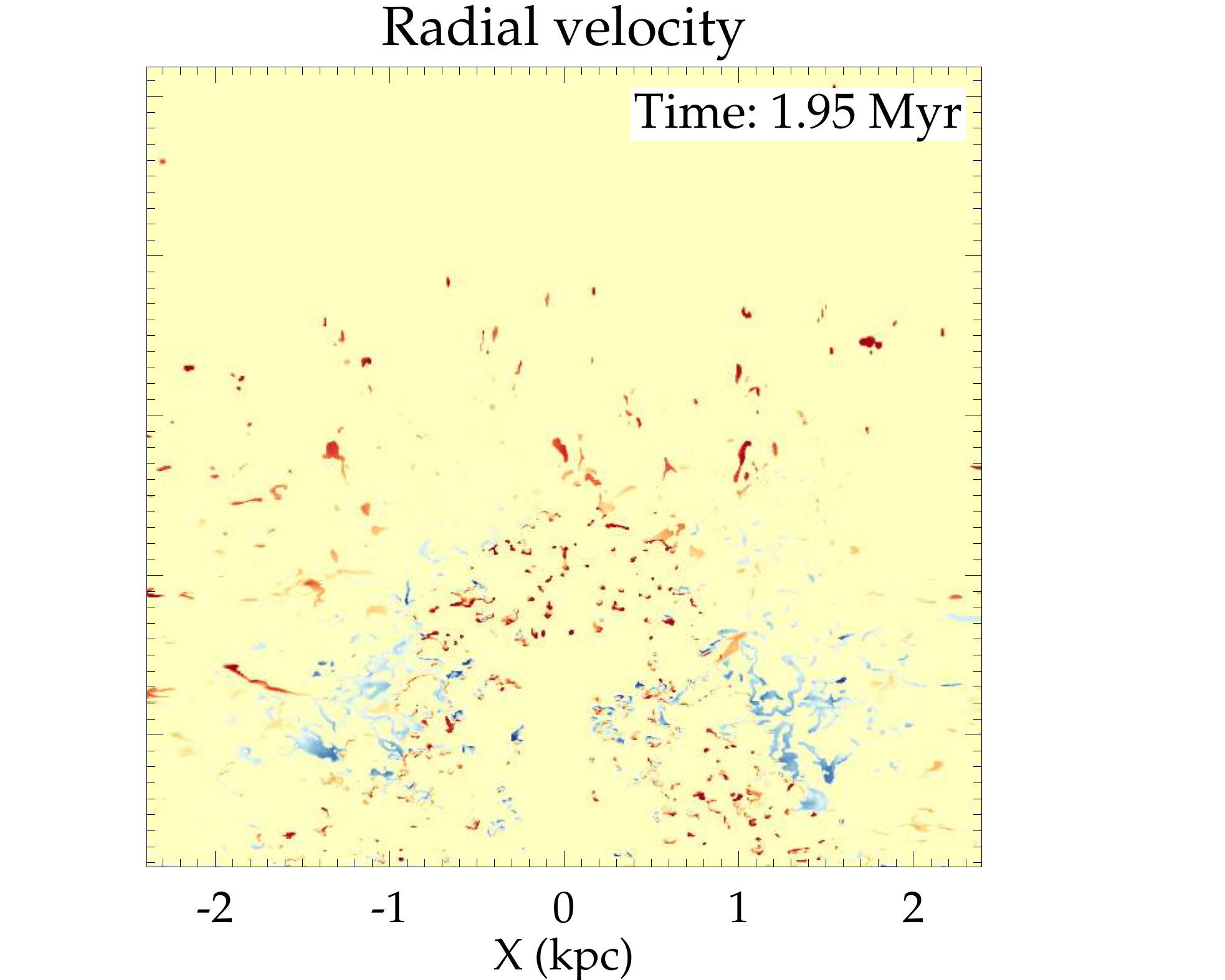}
        \includegraphics[width=5.6cm]{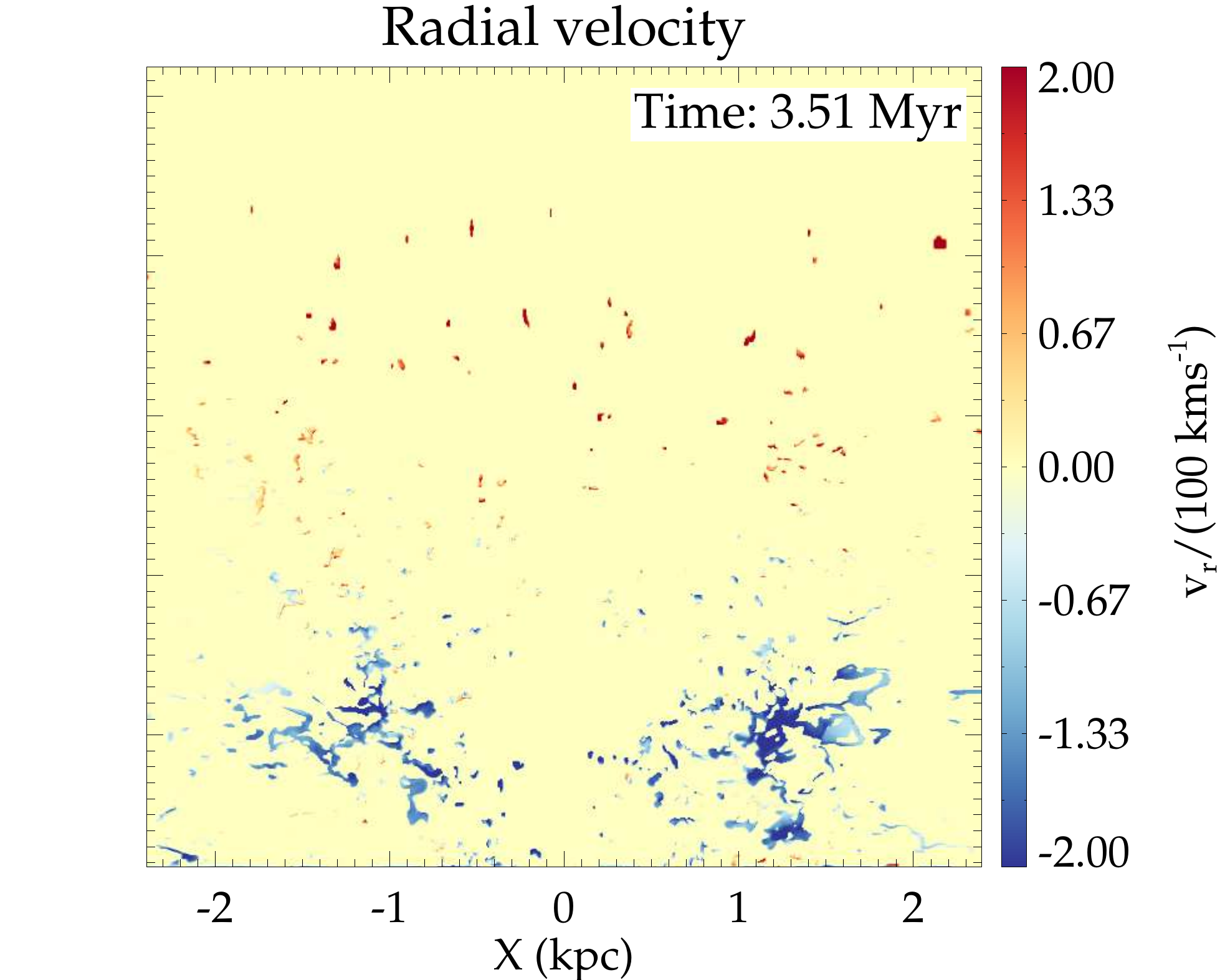}
           }
        }
    \caption{Evolution of density (in units of cm$^{-3}$), temperature (in K) and velocity in units of 100 km s$^{-1}$ in the simulation of propagation of a low-power jet with P$_{\rm jet} = 10^{44}$ ergs s$^{-1}$. The low-power jets remain confined within the ISM for a longer period of time compared with high-power jets, affecting a larger volume of the ISM  \citep{Mukherjee2016,Mukherjee2017}.}
    \label{f:mukherjee_sim}
\end{figure*}

High angular resolution observations at both optical and millimeter wavelengths have provided a wealth of information on jet-ISM interactions, and provided valuable inputs for comparison with the results of numerical simulations. The importance of low-power jets in nearby AGN as an important source of feedback on sub-kpc scales has been highlighted in a number of recent studies. Seeing-limited optical integral field spectroscopic observations from the Multi Unit Spectroscopic Explorer (MUSE) at the Very Large Telescope have been used to study nearby Seyfert galaxies in a survey called Measuring Active Galactic Nuclei Under MUSE Microscope or MAGNUM (\citealt{Venturi2021}, and references therein). The jets in a sample of galaxies studied by \citet{Venturi2021} are $<$1 kpc, have low power ($< 10^{44}$ ergs~s$^{-1}$) and are inclined with 45$^\circ$ to the galaxy disc. They find evidence of enhanced line widths (800 - 1000 km s$^{-1}$) which extended ($>$1 kpc) in directions perpendicular to the jets and the AGN ionisation cones. They interpret this to be due to jet-ISM interactions, showing that these low-power jets are also capable of affecting the host galaxies. A similar result has been
seen in the `radio-quiet' quasar J1316+1753 which have low-power radio jets inclined to the galaxy disk plane \citep{Girdhar2022}. Combining MUSE and ALMA observations \citet{Girdhar2022} report evidence of turbulent gas driven perpendicular to the jet axis, and extending to $\sim$7.5 kpc on opposite sides. They also find evidence of increased stellar velocity dispersion along the jet axis and co-spatial with it, and evidence of both positive and negative feedback. While highly turbulent material appears to escape the galaxy inhibiting star formation, the jets also appear to compress gas in the disk forming new stars which contribute to the stellar bulge. The stellar bulge is closely aligned with the radio jet axis \citep{Girdhar2022}. One of the very striking examples of massive molecular outflow is in the nearby low-luminosity compact radio galaxy B2 0258+35 where about 75 per cent of the central molecular gas is driven outwards by a jet in a radiatively inefficient AGN (Fig.~\ref{f:Murthy_CO_gas}, \citealt{Murthy2022}). 

Outflow of molecular gas and turbulence injected into the ISM will affect star formation in the host galaxy. 
For example, high-resolution observations of the nearby lenticular galaxy NGC1266 harbouring an AGN show that molecular gas is being driven out of the nuclear region at a rate of $\sim$110 M$_\odot$ yr$^{-1}$ \citep{Alatalo2015}. Although only a small fraction may escape the galaxy, the molecular gas that remains is very inefficient in forming stars, with star formation being suppressed by a factor of $\approx$50 compared with normal star-forming galaxies if all the gas is forming stars \citep{Alatalo2015}.
The star formation efficiency in the giant radio galaxy 3C326N appears to be 10 to 50 times lower than normal star forming galaxies \citep{Nesvadba2010}. Similarly the star-formation rate surface densities for J2345$-$0449 appear 30 to 70 times lower than the Kennicutt-Schmidt law of star-forming galaxies \citep{Nesvadba2021}. 
\citet{Lanz2016} observed a sample of 22 radio galaxies which were selected due to the presence of warm molecular gas. They modelled the spectral energy distributions from the ultraviolet to the far infrared and found the star formation rate to be suppressed by a factor of about 3 to 6. In about 25 per cent of the sample the suppression was by a factor of more than 10.
They suggest that this is due to radio jets injecting turbulence into the interstellar medium via shocks. The observational results and trends discussed in this subsection are consistent with the results of numerical simulations of jets interacting with a clumpy ISM \citep[e.g.][]{Sutherland2007,Mukherjee2016,Mukherjee2017,Mukherjee2018a,Mukherjee2018b,Mandal2021}.
The low-power jets($<10^{44}$ ergs s$^{-1}$) remain trapped in the ISM of the host galaxy for a much longer period of time compared with jets of higher power, thereby affecting a larger volume of the ISM (Fig.~\ref{f:mukherjee_sim}; \citealt{Mukherjee2016,Mukherjee2017}). 

\citet{Kalfountzou2017} present far-infrared 
observations with Herschel of 74 radio-loud quasars, 72 radio-quiet quasars and 27 radio
galaxies (RGs) over the redshift range $0.9<z<1.1$
and investigate the dependence of star formation rate 
on AGN luminosity, radio loudness and orientation. They suggest that there is a jet power threshold where feedback switches from compressing gas and enhancing star formation to heating and ejecting gas and thereby suppressing star formation. Both observational and theoretical work on jet-cloud interactions and perhaps a deeper understanding of star formation itself will enhance our understanding of these aspects.

\subsection{Jet feedback and x-ray cavities}
\citet{Pedlar1990} observed the FRI radio galaxy NGC1275 in the Perseus cluster over a wide range of frequencies lower than about a GHz, and were among the early ones to highlight the importance of jet feedback
while considering models of cooling flows in clusters of galaxies \citep[e.g.][]{Fabian1981}. Over the years high-resolution x-ray observations have revealed giant x-ray cavities and shock fronts which are closely related to the radio emission in many clusters of galaxies, underlining the importance of radio jet feedback \citep[e.g.][]{McNamara2007,McNamara2012}. Multiple generations of these cavities are signs of episodic AGN activity \citep[e.g.][]{Vantyghem2014}. These cavities also provide a direct and reasonably reliable means of estimating energy injected into the atmospheres by jets in AGN \citep[e.g.][]{Hardcastle2020}.

\section{Recurrent jet activity}
\begin{figure}[ht!]
    \centering
    \includegraphics[width=8.0cm]{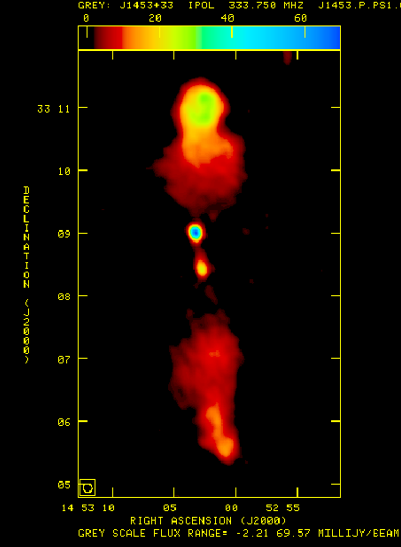}
    \caption{GMRT image of the double-double radio galaxy J1453+3308 at 330 MHz \citet{Konar2006}.}
    \label{f:J1453_ddrg_Konar}
\end{figure}

Radio galaxies have been found to show evidence of episodic or recurrent nuclear activity since the 1980s. For example, a radio jet south of the nucleus in the radio galaxy 3C338 has been suggested to be due to an earlier cycle of jet activity \citep{Burns1983}. Sharp discontinuities in the spectral index distributions of the lobes, where emission from the earlier cycle of activity have a significantly steeper spectral index, are signs of recurrent jet activity. Examples of such sources include 3C388 \citep{Roettiger1994,Brienza2020} and Her A \citep{Gizani2005}. Old electrons from an earlier cycle of activity could scatter low-energy ambient photons to high energies in the x-ray region of the spectrum via inverse-Compton scattering. \citet{Steenbrugge2008} have suggested an earlier cycle of activity in the archetypal FRII radio galaxy Cygnus A from x-ray observations. In the extreme case, an old radio galaxy may
be visible only at x-ray wavelengths due to inverse-Compton scattering of the ambient photons. One such example is inverse-Compton ghost of a giant radio source HDF130 in the Hubble Deep Field, where low-frequency observations with the GMRT did not reveal any radio emission \citep{Mocz2011b}. LOFAR observations of radio galaxies at low frequencies 
show a variety of signatures of recurrent jet activity
\citep{Jurlin2020,Shabala2020}. The most striking examples of episodic jet activity are the double-double radio galaxies of DDRGs (Fig.~\ref{f:J1453_ddrg_Konar}) which have two pairs of radio lobes on opposite sides of the parent optical object \citep{Schoenmakers1999,Saikia2009,Kuzmicz2017}. In a couple of
cases three pairs of radio lobes indicating three cycles of jet activity have been seen \citep{Brocksopp2007,Hota2011}.

\citet{Kuzmicz2017} compiled a sample of 74 extragalactic radio sources with evidence of recurrent jet activity, of which 67 are galaxies, 2 are quasars, and 5
are unidentified sources. They found the black hole masses of rejuvenated radio sources and a control sample of FRII sources to be similar. However they find a difference in optical morphology, which they interpret to be due to merger events in the history of the host galaxy of restarted radio sources. From a rather small sample of sources \citet{Saikia2009} and \citet{Chandola2010} suggested a higher incidence of H{\sc i}
absorption towards the nuclear regions of rejuvenated radio sources. Any differences between rejuvenated radio sources and control samples with evidence of a single cycle of activity needs further investigation using larger samples. The number of rejuvenated radio galaxies and quasars will increase with more sensitive observations especially at low frequencies as has been demonstrated by LOFAR observations of the HETDEX field and the Lockman Hole region \citep{Mahatma2019,Jurlin2020}.

The time scale of recurrent activity is likely to have a wide range. \citet{ODea2021} list seventeen CSS/PS sources with evidence of diffuse lobes of emission from an earlier cycle of activity. Several of these appear to have diffuse emission on only one side of the active nucleus; the one on the opposite side possibly being below the detection threshold. \citet{Stanghellini2015} estimated these relics to be from about $10^7$ - $10^8$ yr ago. Spectral and dynamical age estimates as well as statistical studies suggest similar time scales of the  jet activity \citep{Konar2006,Shulevski2012,Shabala2008,Best2005}, although there are suggestions of smaller time scales as well, as in the case of 3C293 \citep{Joshi2011} and CTA21 \citep{Salter2010}. \citet{Reynolds1997} suggested that jets in CSS/PS sources may be intermittent on time scales of $\sim$10$^4$ - $10^5$ yr. The physical processes responsible for recurrent jet activity remains to be understood and may also provide insights towards understanding the triggering of the powerful radio jets in radio-loud AGN.

\section{Concluding remarks}
Sensitive, high-resolution observations at different wavelengths across the electromagnetic spectrum and monitoring programs, along with theoretical modelling and numerical simulations over the last decade or so have significantly enhanced our understanding of jets in AGN. However many of the fundamental questions related to jet physics such as jet launching and collimation,
jet composition, magnetic fields, particle acceleration
and constituents on different scales, remain largely unanswered. At radio frequencies the advent of the Square Kilometer Array (SKA) with unprecedented sensitivity and resolution in both total intensity and polarization, along with SKA1-VLBI, is likely to have a huge impact in our understanding of jets in AGN (cf. \citealt{Laing2015,Agudo2015}). 
Here we summarise a few aspects where we are likely to see significant advances over the next decade or so.

Although jets are seen over a wide range of luminosities and different host galaxies, what determines the launching of jets of different jet powers? This is also related to our understanding of the radio-loud radio-quiet dichotomy, although the distribution of the radio loudness parameter may not be strongly bimodal as was once seen. There may be a smooth transition from the radio quiet to the radio loud regime (\citealt{Macfarlane2021} and references therein). An early study of the Palomar-Green quasar sample suggested that almost all quasars with the mass of the SMBH M${_\bullet} > 10^9$ M$_\odot$ are radio loud while those with M${_\bullet} < 3\times10^8$ M$_\odot$ tend to be radio quiet \citep{Laor2000}. But there are  radio-quiet as well as radio loud objects in the range M${_\bullet} > 3\times10^8$ M$_\odot$ and  M${_\bullet} < 10^9$ M$_\odot$, suggesting that mass alone may not be adequate to understand the radio loud - radio quiet dichotomy or the launching of luminous radio jets. The other parameter to consider is black hole spin \citep[e.g.][]{Chiaberge2011}. However, spin alone also may not be a sufficient condition for launching luminous radio jets as rapidly spinning SMBHs have been seen in both radio-loud and radio-quiet objects although it may be a necessary condition \citep[cf.][]{Reynolds2019}. Comparing the sources from LOFAR Two-Metre Sky Survey (LoTSS) DR1 with the Sloan Digital Sky Survey (SDSS) DR7, \citet{Sabater2019} find AGN activity to show a strong dependence on both stellar and black hole masses with massive galaxies above $10^{11}$ M$_\odot$ almost invariably exhibiting radio AGN activity. This is perhaps not surprising considering the Magorrian relationship between black hole and galactic bulge masses \citep{Magorrian1998}. The fundamental plane for black holes in the x-ray band \citep{Merloni2003} and also in the optical band \citep{SaikiaP2015} illustrate the close relationship between radio luminosity, black hole mass and x-ray/optical line luminosity. This has recently been extended to incorporate the spin of the black hole for a sample of flat-spectrum radio quasars and BL Lac objects \citep{Chen2021}. Perhaps stellar and black hole mass and spin, and the accretion process coupled with the availability of fuel all influence the launching of jets of different powers. On the theoretical front magnetic fields appear to play an important role \citep{Blandford1977,Blandford1982}, and this has been explored using fully 3D GRMHRD simulations \citep[e.g.][]{McKinney2012}.

What is the composition of the AGN jets and how do they change with distance from the central engine? In the initial acceleration phase jets are believed to be Poynting flux dominated, later converting to particle dominated plasma. The point of conversion is unclear and requires an understanding of the particle acceleration processes (see \citealt{Agudo2015} for a discussion). There is increasing evidence that jets consist of an inner spine and an outer layer. Is the constitution of the inner spine and the outer layer the same? Could the inner spine be made of an electron-positron plasma and the outer an electron-proton plasma as the jets propagate in the vicinity of the accretion disk and black hole? How does the constitution change with distance as jets propagate outwards for the FRI and FRII sources? Polarization observations especially circular polarization observations will provide important inputs towards understanding the constitution of jets.

Other important parameters related to the jets are the velocity fields, magnetic field structure and jet power and the environment.
For the low-luminosity FRI Laing \& Bridle (see Section 7.1) have constructed detailed models and showed that the velocity decreases with distance from the 
central engine. \citet{Laing2014} find the magnetic field component to be predominantly longitudinal close to the
AGN and toroidal after recollimation.
Increased resolution and sensitivity will enable such studies to be extended to FRII sources \citep{Laing2015}. Detailed polarization studies will enable estimates of the Faraday depth and its variation in the jets, estimate the thermal particle content and explore whether the confinement of jets is due to magnetic fields or thermal pressure if they are confined. VLBI-scale observations suggest toroidal fields, but increased sensitivity and resolution will enable this to be explored over a range of length scales for large samples.

What is the source of high-energy emission from AGN jets?
Observations with the Chandra telescope demonstrated the ubiquity of AGN jets at X-ray wavelengths with individual knots of emission being detected at X-ray wavelengths. 
For FRI sources the x-ray emission from the jets appear to be due to synchrotron radiation, while for FRII jets 
inverse-Compton radiation from jets moving at highly relativistic velocities with Lorentz factors of 10 - 20 has been the standard explanation. An alternative explanation is that the high-energy emission is from a second electron population, which has been suggested for example from studies of the quasar 3C273 and other sources \citep[e.g.][]{Meyer2014,Meyer2015}. Such multiwavelength studies which identify sources of high-energy emission and also the location within the source responsible for the emission are important to understand the physical processes in the jets. In the context of high-energy emission from radio sources, it is relevant to note that there is a growing population of radio galaxies which are $\gamma$-ray sources. \citet{Bruni2022} have reported a new $\gamma$-ray emitting FRII radio galaxy which they model as arising from the radio lobes due to inverse-Compton scattering of the photons off the radiating electrons. They have listed another 8 such radio galaxies.

Are all AGN jet activity episodic? Drawing an analogy from microquasars, \citet{Nipoti2005} suggested that radio loudness may only be a function of epoch, and there may be no essential difference between radio-loud and radio-quiet objects. More recently \citet{Moravec2022} have explored whether different kinds of AGN reflect x-ray binary spectral states, and find that ``radio-loud AGN occupy distinct areas of the hardness-intensity diagram depending on the morphology and excitation class, showing strong similarities to x-ray binaries''. Although these are interesting approaches, a deeper understanding of how a powerful radio jet is launched may help clarify whether all AGN are episodic. On the observational side LOFAR observations have revealed many more sources with signs of recurrent activity. More sensitive observations with SKA is also likely to reveal how common is the evidence for earlier cycles of activity. A deep search with the GMRT showed that these are relatively rare \citep{Sirothia2009a}, but deeper observations are required. Besides radio observations deep x-ray surveys may also reveal many double-lobed sources which are too weak to be seen at radio wavelengths but may be visible at x-ray wavelengths due to inverse-Compton scattering of ambient photons by the low-energy electrons (see \citealt{Mocz2011b} for an example). An understanding of the frequency of recurrent AGN jet activity and their duty cycles are vital for understanding a number of aspects related to AGN feedback, including the evolution of galaxies.

In the last few years, new telescopes or upgraded versions of earlier ones have yielded many interesting results some of which have been highlighted in this short review. JWST, which has been launched, and upcoming telescopes such as SKA, TMT and ATHENA, to name a few, should yield a wealth of information to help answer many of outstanding questions related to AGN jets.

\section*{Acknowledgments}
It is a pleasure to thank Alan Bridle and Dipanjan Mukherjee, who was the reviewer, Pratik Dabhade and Mousumi Mahato for their detailed and helpful comments on the manuscript. I am extremely grateful to Bia Boccardi, Alan Bridle, Jim Condon, Bill Cotton, Bernd Husemann, Robert Laing, Beatriz Mingo, Dipanjan Mukherjee, Suma Murthy, Hiroki Okino, Alice Pasetto and Peter Thomasson, for kindly providing the figures and also for their permission to reproduce the figures. Thanks also to the Editor-in-chief, Astronomy and Astrophysics, and American Astronomical Society for permission to reproduce figures. I am also very grateful to Pratik Dabhade for his generous help in getting the figures and the references organized for the JoAA format. I also wish to express my gratitude to the organizers of the ARIES conference on jets titled Astrophysical jets and observational facilities: National perspective, and the editors of the proceedings, Shashi Pandey, Alok Gupta and Sachindra Naik, for asking me to write an extended version based of my talk and above all, waiting patiently for it in spite of a long delay on my part.

\end{document}